\documentclass{aa}  

\usepackage{graphicx}
\usepackage{txfonts}
\usepackage{color} 

\newcommand{\kB}{k_\mathrm{B}}

\begin{document}

   \title{{\sl Ab initio} based equation of state of dense water for
planetary and exoplanetary modeling}

   \subtitle{}

   \author{
     S. Mazevet
          \inst{1,2}
     \and
     A. Licari\inst{1,3}
     \fnmsep\thanks{Current address:
      Lyc\'ee Jean Dautet, La Rochelle, France}
          \and
          G. Chabrier\inst{3,4}
          \and
          A. Y. Potekhin\inst{5}}

   \institute{Laboratoire Univers et Theories, Universit\'e Paris Diderot,
Observatoire de Paris, PSL University, 5 Place Jules Janssen, 92195 Meudon
France\\
\email{stephane.mazevet@obspm.fr}
     \and
     CEA-DAM-DIF, 91280 Bruyeres Le Chatels, France
     \and
     CRAL, Ecole Normale Sup\'erieure de Lyon, UMR CNRS 5574, All\'e d'Italie, Lyon,  France
     \and
School of Physics, University of Exeter, Exeter, UK EX4 4QL
\and
      Ioffe Institute, Politekhnicheskaya 26, 
      St. Petersburg 194021, Russia}

   \date{A\&A \textbf{621}, A128 (2019);
   \textbf{\textit{Eq.\,(13) and the value of $S_0$
   are corrected after the publication}}}

% \abstract{}{}{}{}{} 
% 5 {} token are mandatory
 
  \abstract
      {The modeling of planetary interiors requires accurate equations of
        state for the basic constituents with proven validity in the
        difficult pressure-temperature regime extending up to 50,000~K and
        hundreds of Mbars. While equations of state based on first
        principles simulations are now available for the two most abundant
        elements, hydrogen and helium, the situation is less satisfactory
        for water where no wide-range equation of state is available despite its
        need for interior modeling of planets ranging from super-Earths to
        planets several times the size of Jupiter. }
      {    
        As a first step toward a multi-phase equation of state for dense
        water, we develop a temperature-dependent equation of state for
        dense water covering the liquid and plasma regimes and extending
        to the super-ionic and gas regimes. This equation of state covers
        the complete range of conditions encountered in planetary
        modeling.
      }
      {We use first principles quantum molecular dynamics simulations and
        its Thomas-Fermi extension to reach the highest pressures
        encountered in giant planets several times the size of Jupiter.
        Using these results, as well as the data available at lower
        pressures, we obtain a parametrization of the Helmholtz free
        energy adjusted over this extended temperature and pressure
        domain.  The parametrization ignores the entropy and density jumps
        at phase boundaries but we show that it is sufficiently accurate to model
        interior properties of most planets and exoplanets.  }
      % results heading
      {We produce an equation of state  given in analytical
        form that is readily usable in planetary modeling codes
        and dynamical simulations (a fortran implementation can be found at http://www.ioffe.ru/astro/H2O/).
        The EOS produced is valid for the entire density range
        relevant to planetary modeling, 
        for densities where quantum effects for the ions can be neglected, and for temperatures below 50,000K.
        We use this equation of state to
        calculate the mass-radius relationship of exoplanets up to
        5\,000$M_\mathrm{Earth}$, explore temperature effects in ocean and
        wet Earth-like planets, and quantify the influence of the water
        EOS for the core on the gravitational moments of Jupiter.}
   {}


   \maketitle
%
%-------------------------------------------------------------------

\section{Introduction}

With the advent of a new generation of space and ground based instruments,
the constraints of the interior structure of planets and exoplanets have
been continuously improving. This is for example the case with Jupiter
where the Juno space mission \citep{Bolton_etal17} is currently measuring
gravitational moments to an unprecedented accuracy, or the various transit
and radial velocity programs such as HARPS or Kepler that provide, when
combined, density measurements for more than 600 exoplanets
\citep{exoplanet.eu}. These continuously improving observational
constraints on the inner structure of planets and exoplanets call for a
proportional effort on the modeling side to achieve a better understanding on
the nature of these objects. The modeling of planetary interiors directly
relies on the thermodynamic properties of matter at the extreme
temperature and pressure conditions encountered within a planet. These can
reach several hundreds of Mbars (1~Mbar=100~GPa) and up to 500\,000~K for
the brown dwarfs and giant planets several times the size of Jupiter 
\citep[see, e.g.,][for review]{Baraffe_Chabrier_Barman10}.

Great progress has been made over the past ten years at understanding this
extreme state of matter, which is not directly accessible to laboratory
experiments, by using first principles or \textit{ab initio} simulations
based on density functional theory \citep{Benuzzi-Mounaix_etal14}. This
computational intensive approach, that can be validated on the limited
density-temperature range accessible to shock or high-pressure
experiments, provides a fully quantum mechanical description for the electronic structure of
this state of matter without adjustable parameters. With computational resources
greatly increasing, this approach provides the most reliable means to
calculate the properties of matter in the thermodynamical range most
relevant to planetary modeling, extending from the experimentally
accessible thermodynamic conditions to the ones where analytical and
semi-analytical approaches become valid. This method has been recently applied
to provide comprehensive equations of state (EOS) for the two most
abundant elements, hydrogen and helium
\citep{Caillabet_Mazevet_Loubeyre11,Becker_etal14,Militzer13}, which brought
about a renewed understanding of the internal structure of Jupiter

\citep{Nettelmann_etal12,Hubbard_Militzer16,Militzer_etal16,Wahl_etal17,Guillot_etal18}.
A similar effort is underway for water, and \textit{ab initio} simulations
are now probing the physical properties of water at conditions encountered
in planetary interiors.

Following the pioneering work of \citet{Cavazzoni_etal99},
\citet{Mattsson_Desjarlais06} and \citet{French_etal09} calculated the
properties of the superionic phase for dense water at conditions
encountered within Uranus and Neptune. For water, the superionic phase is
defined as oxygen atoms locked into either a body-centered cubic (BCC) or
face-centered cubic (FCC) crystalline structure with the hydrogen atoms
diffusing like in a liquid. This particular phase, which appears at
pressures and temperatures above the regular solid ice phases, 
provides electrical conductivities
compatible with the unusual magnetic fields observed for these
objects \citep{Redmer_etal11}. Subsequent works attempted to identify the
stable solid phase underlying the superionic region of the phase diagram
\citep{Wilson_Wong_Militzer13,French_Desjarlais_Redmer16} and investigated
the miscibility of water in a H-He dense plasma anticipated near Jupiter's
core \citep{Soubiran_Militzer15}. While the debate is ongoing regarding
the precise localization and nature of the superionic phase for dense
water \citep{Millot_etal18}, there is still no comprehensive EOS of dense
water available for planetary modeling. 

As the focus in exoplanetary science is now turning to characterize the
Earth-like to Neptune-like continuum, there is a great need for an EOS for
water that spans thermodynamic conditions ranging from the atmosphere of
an Earth-like planet to the core of a giant planet or brown dwarf several
times the size of Jupiter. In the first part of the manuscript, we expand
on the work of \citet{French_etal09} and apply \textit{ab initio}
molecular dynamics simulations and its high-pressure high-temperature
Thomas Fermi limit to calculate the properties of water up to a density of
100 g~cm$^{-3}$ and reach conditions encountered in these massive objects.
We supplement this data set by the free-energy  parametrization developed
by the International Association for Properties of Water and Steam
(IAPWS)\footnote{\url{http://www.iapws.org}} \citep{Wagner_Pruss02} that
provides an accurate account of the behavior of water in the vapor and
liquid phases at pressures below 1~GPa. Using these data sets, we 
built a wide-range EOS that covers the complete thermodynamical state relevant for planetary modeling.
We approximate this EOS by an analytical fit of the free energy, whose derivatives
simultaneously provide the fits to pressure and internal energy in
agreement with the data, and provides an estimation for the total entropy.

We apply this EOS to probe the effect
of temperature on the standard mass-radius diagram used to identify
exoplanets by considering wet Earth-like, ocean planets and pure water
planets. Finally, we use this EOS for dense water to calculate the
gravitational moments of Jupiter currently measured by the Juno probe
\citep{Guillot_etal18}.

%--------------------------------------------------------------------
\section{\textit{Ab initio} simulations}
\label{sect2}

The EOS developed in the present work is based on \textit{ab initio}
molecular dynamics simulations for densities $\rho$ between 1 g~cm$^{-3}$
and 50 g~cm$^{-3}$ and temperatures $T$ between 1000~K and 50\,000~K,
complemented with the free-energy parametrization of
\citet{Wagner_Pruss02} at lower densities and temperatures. At $\rho > 50$
g~cm$^{-3}$, we used the Thomas-Fermi molecular dynamics (TFMD) simulations
\citep{Lambert_Clerouin_Zerah06,Mazevet_etal07}.

%--------------------------------------------------------------------
\subsection{Computational details}

To complement the data obtained previously for water using \textit{ab
initio} molecular dynamics simulations
\citep{French_etal09,Wilson_Wong_Militzer13,French_Desjarlais_Redmer16},
we carried out simulations using the ABINIT \citep{Gonze_etal09}
electronic structure package. This consists in treating the electrons
quantum mechanically using finite temperature density functional theory
(DFT) while propagating the ions classically on the resulting
Born-Oppenheimer surface by solving the Newton equations of motion. We
used the generalized gradient approximation (GGA) formulation of the DFT
with the parametrization of the exchange-correlation functional provided
by  \citet{Perdew_Burke_Wang96} (PBE).

We used two sets of pseudopotentials to cover the density range from $1$
g~cm$^{-3}$ to $50$ g~cm$^{-3}$. For densities up to $5$ g~cm$^{-3}$, we
used two  projector augmented wave (PAW) pseudopotentials generated by
\citet{Jollet_Torrent_Holzwarth14}. These pseudopotentials are designed
to reproduce accurately the all-electrons results obtained for the individual
atomic species. This provides a warranty that the use of the pseudopotential does
not cause any important spurious effect.
For the two atomic species considered here, hydrogen and oxygen, this consists in a
cutoff radius of respectively $0.7a_0$ and $1.2a_0$, 
where $a_0 = \hbar^2/(m_\mathrm{e} e^2)$ is the Bohr radius,
 and an oxygen pseudopotential with the $1$s state treated as a core state. To reach
densities above 7 g~cm$^{-3}$, we use the ATOMPAW
\citep{Holzwarth_Tackett_Matthews01} package to generate pseudopotentials
with cutoff radius of $r_\mathrm{paw}=0.4$ $a_0$
and $r_\mathrm{paw}=0.6$ $a_0$ for, respectively, the hydrogen and oxygen atomic species.
We further find that the oxygen $1$s state needs to be included as a valence state to
reach the highest density treated, $50$ g~cm$^{-3}$. The accuracy of the
two pseudopotentials produced was inferred by directly comparing the cold
curves obtained for the individual atomic species in the FCC
phase with the corresponding all-electrons calculations \citep{Jollet_Torrent_Holzwarth14}.
This significant reduction in the cutoff radius requires an increase of the plane wave cutoff
from 30 to 100~Ha to reach convergence in pressure and energy below 1\%.

The convergence tests performed regarding the number of particles in the simulation
cell and the {\bf k}-point grids in momentum space confirm the results reported by
\citet{French_etal09,French_Desjarlais_Redmer16}.  We paid particular
attention to the influence of the superionic phase and performed
calculations using both the FCC and BCC crystallographic structures

\citealt{Wilson_Wong_Militzer13} pointed out that a superionic phase where
the oxygen ions remain in an FCC rather than BCC structure may be more
stable at intermediate temperature. For the EOS points, we used 54 atoms in
the BCC superionic phase. For the FCC superionic phase, we used 108 atoms
for densities below and up to 15 g~cm$^{-3}$ while we found sufficient to
use 32 atoms at the highest densities. For both phases, we performed the
simulations at the $\Gamma$-point and integrated the equations of motion
with a time-step of $5$~a.u. ($1$~a.u.\,=\,0.024~fs). We attribute this
slight difference with the simulation parameters reported by 
\citet{French_Desjarlais_Redmer16} to the level of accuracy required in
their calculations to evaluate the thermodynamic potentials in the
superionic FCC and BCC phases.

\subsection{\textit{Ab initio simulation results}}

\begin{figure}[t]
   \centering
   \includegraphics[width=\columnwidth]{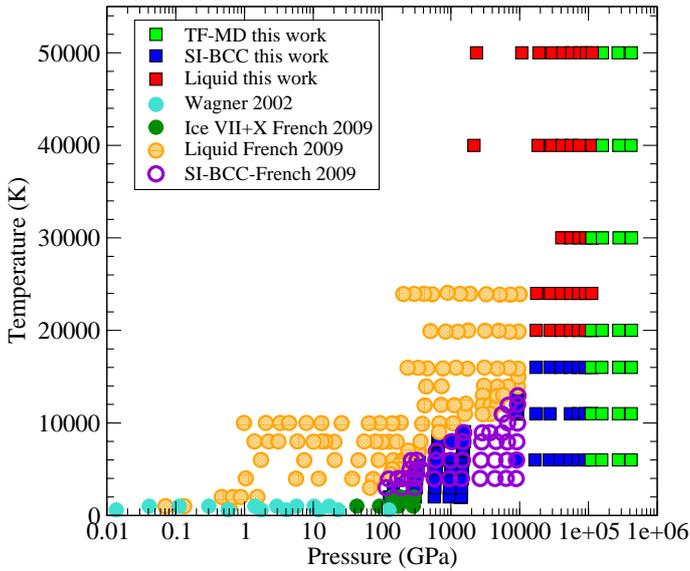}
   \caption{ The phase diagram of dense water obtained using the
    \textit{ab initio} and Thomas-Fermi simulations. Each symbol
    represents a simulation point. The phase state is indicated by a
    colored symbol according to the legend. Previous \textit{ab initio}
    points obtained by  \citet{French_etal09} as well as representative
    points of  \citet{Wagner_Pruss02} are also shown. }
              \label{fig1}%
  \end{figure}

Figure~\ref{fig1} displays the pressure $P$ and temperature $T$ values at which
\textit{ab initio} simulations were performed.  We have also included
sample points from the IAPWS free energy formulation 
\citep{Wagner_Pruss02} for completeness. For the internal energy and
pressure, we found a good agreement between our calculations  and the
previous results of \citet{French_etal09}. We have thus directly included these
points in our \textit{ab initio} set. Figure~\ref{fig1} also shows  that a superionic
phase remains stable up to the highest pressures for temperatures up to 16000~K. 

The superionic phase is identified in our molecular dynamics simulations
by looking at the mean square displacement of the hydrogen and oxygen ions
as a function of time. Figure~\ref{fig1} shows that at $\rho > 15$
g~cm$^{-3}$ the superionic phase remains stable up to $T=16000$~K
when considering either the BCC or FCC structures. With the temperature
grid used here, this suggests that the superionic-plasma phase boundaries
for both the BCC and FCC structures vary slowly in this pressure range and
are both located between 16000~K and 20000~K. We also point that the
simulations performed here do not allow us to identify the superionic
phase that is the most stable in this thermodynamical regime. They do not
indicate either whether another superionic phase may exist in this
thermodynamical range.     

Here, we do not explore further the exact determination of either the
superionic-plasma boundary nor the nature of the superionic state. The
results previously obtained at low pressures indicated that this issue has
little consequences for the EOS \citep{French_Desjarlais_Redmer16}. To
confirm that this remains the case for the entire density range considered
here, we show in Table~\ref{tab1} the results obtained for the internal
energy and pressure at representative densities and considering both the
BCC and FCC superionic states. We see in Table~\ref{tab1} that the
pressure and internal energy values agree to within 1.5\%. We thus started
our simulations with the BCC superionic lattice at $\rho > 20$
g~cm$^{-3}$, as this enables smaller simulation cells. We tested using
larger simulation cells that convergence is reached for pressure and
internal energy up to 50 g~cm$^{-3}$.  

\begin{table*}[ht]
\centering
\begin{tabular}{*{8}{c}}
    \hline
   \hline
$\rho$~(g~cm$^{-3}$)&$T$~(K)&$P_\mathrm{BCC}$~(Mbar)&$\Delta P$~(Mbar)&${\Delta
P}/P$&$U_\mathrm{BCC}$~(eV/atom)&${\Delta U}$~(eV/atom) & $|\Delta U/U|$\\
    \hline
    7  & 6000  &  14.85 &  0.22 & 1.5\% & $-685.9$  & 0.03  & 0.004\%\\
    15 & 6000  &  90.14 &  1.13 & 1.3\% & $-668.7$  & 0.52  & 0.08\%  \\
    20 & 6000  & 166.80 &  1.47 & 0.9\% & $-656.5$  & 0.66  & 0.10\%  \\
    25 & 6000  & 267.63 &  1.91 & 0.7\% & $-643.5$  & 0.76  & 0.12\%  \\
    40 & 11000 & 709.97 &  0.9  & 0.13\%& $-601  $  & 1.0   & 0.17\%  \\    
\hline
\end{tabular}
\caption{Pressure $P_\mathrm{BCC}$ and internal energy $U_\mathrm{BCC}$
obtained for the BCC superionic phase and the differences $\Delta P =
P_\mathrm{FCC}-P_\mathrm{BCC}$ and $\Delta U =
U_\mathrm{FCC}-U_\mathrm{BCC}$ between the FCC and BCC superionic
phases, as well as fractional differences. 
The reference energy corresponds to the total
binding energy of the ground state
of an isolated water molecule ($693.3$ eV/atom).
\label{tab1}
}
\end{table*}

\subsection{Thomas-Fermi extension}

\begin{figure}
   \centering
   \includegraphics[width=.8\columnwidth]{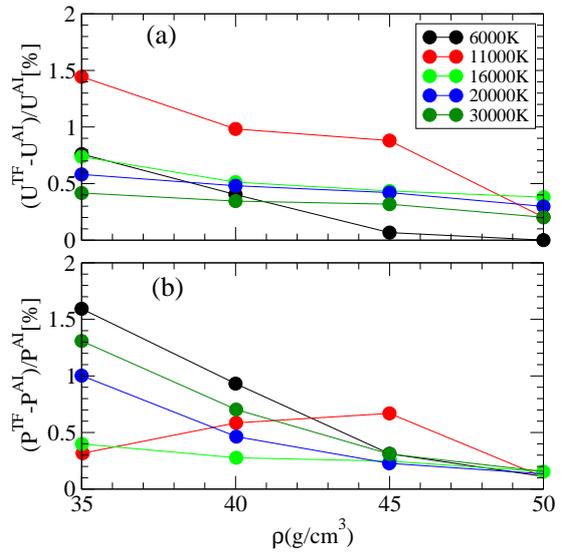}
   \caption{(a) Relative difference between the \textit{ab initio} and Thomas-Fermi internal energies as a function of density. 
   (b) Same as in (a) but for the pressures.}
              \label{fig2}%
  \end{figure}

Beyond 50 g~cm$^{-3}$, we switch from full \textit{ab initio} simulations
to Thomas Fermi molecular dynamics simulations (TFMD) \citep{Lambert_Clerouin_Zerah06}.
This consists in using the Thomas-Fermi approximation to describe the electrons while propagating the
ions on the resulting Born-Oppenheimer surface. In this framework, the
kinetic energy operator in the electronic Hamiltonian is replaced by a
functional of the density \citep{martins}. This greatly simplifies the
calculation as a plane wave basis is no longer needed and the electronic
density is obtained by just solving the Poisson equation. This represents
the natural high-density limit to the DFT and hence to the \textit{ab
initio} simulations. We show in Fig.~\ref{fig2} the relative difference
for the pressure $P$ and internal energy $U$ between the full \textit{ab
initio} and the TFMD simulations at densities between 34 and 50
g~cm$^{-3}$. For both quantities, we see that the difference between the
two methods is rapidly reducing as the density increases to be well under
1\% at 50 g~cm$^{-3}$. This result clearly shows that the high-density
limit is reached and justifies the use of the Thomas-Fermi approximation beyond this density. 

  We further point out that the Thomas-Fermi approximation requires the
use of a regularization potential. We make the choice of using the
pseudopotential determined at $\rho=75$ g~cm$^{-3}$ and $T=2000$~K
throughout the entire range of interest. This approximation introduces an
uncertainty on the internal energy of a few percent as the regularization
formally breaks the transferability of the pseudopotential in density and
temperature \citep{Lambert_Clerouin_Zerah06}. The internal energy obtained
by the TFMD method is adjusted  to the \textit{ab initio} one at $\rho=50$
g~cm$^{-3}$ and $T=6000$~K. While the simulations were all started in the
FCC phase, we make the choice of not recording the stability of the
superionic phase as quantum effects for the protons may start to play a
non-negligible role in its stability \citep{French_Desjarlais_Redmer16}.
The effect on the pressure and internal energy is a higher order
effect.

  \section{Analytical fit of the Helmholtz free energy}
  \label{sect:fit}
  
The 
full \emph{ab initio}
simulation results presented in the previous section are used to
construct a functional form of the Helmholtz free energy valid over the
entire density-temperature domain relevant to planetary modeling. Such an
analytical fit of the free energy  provides a convenient means to combine
various data for use in simulations of interior structure or evolution of
planets. For water, this includes the \textit{ab initio} results presented
above, valid down to about a few GPa, and the low-energy IAPWS formulation
\citep{Wagner_Pruss02} that provides the EOS of water at
$P<1$~GPa in the vapor and liquid phases as constrained by
experimental measurements.

\subsection{Formulation}

The parametrization of the Helmholtz free energy 
is expressed as
\begin{equation}
F =F_\mathrm{tran}+w(\rho,T) F_\mathrm{low}
     +[1-w(\rho,T)] F_\mathrm{high}
     +F_T-S_0T,
\label{fit}
\end{equation}
where each term has its own physical meaning.

The first term
\begin{equation}
  F_\mathrm{tran} = N_{\mathrm{H}_2\mathrm{O}}\kB T
   \left[\ln(n_{\mathrm{H}_2\mathrm{O}}\lambda_{\mathrm{H}_2\mathrm{O}}^3)-1
    \right]
\label{Ftran}
\end{equation}
is the translational (ideal molecular gas) contribution. Here,
$N_{\mathrm{H}_2\mathrm{O}}=n_{\mathrm{H}_2\mathrm{O}}V
=N_\mathrm{at}/3$
is the number of H$_2$O molecules,  $N_\mathrm{at} = n_\mathrm{at} V =
3\rho V/m_{\mathrm{H}_2\mathrm{O}}$ is the total number of H and O atomic
nuclei in volume $V$,
$m_{\mathrm{H}_2\mathrm{O}}=
2m_\mathrm{H}+m_\mathrm{O}=2.99\times10^{-23}$~g is the mass of the
molecule,
$\kB$ is the Boltzmann constant.
and
\begin{equation}
   \lambda_{\mathrm{H}_2\mathrm{O}} = 
    \left(\frac{2\pi\hbar^2}{ m_{\mathrm{H}_2\mathrm{O}}
     \kB T}\right)^{1/2}
\end{equation}
is the thermal wavelength of a molecule.

In the second and third terms of Eq.~(\ref{fit}),
\begin{eqnarray}
   F_\mathrm{low} &=& \frac{N_{\mathrm{H}_2\mathrm{O}}^2}{V}\,
          ( b_\mathrm{vdW} \kB T - a_\mathrm{vdW} )
\nonumber\\& + &
           \frac23
          N_{\mathrm{H}_2\mathrm{O}} \kB T\, (b_\mathrm{vdW} n_{\mathrm{H}_2\mathrm{O}})^{3/2} 
           [1+(390.92\mbox{\,K}/ T)^{2.384}]
\label{Flow}
\end{eqnarray}
and
\begin{equation}
   F_\mathrm{high}(N_\mathrm{at},T,V) =
    F_\mathrm{e}(N_\mathrm{at} Z_*,T,V)
\label{Fpl}
\end{equation}
are the analytical expressions for the excess free energy in the
moderate-density liquid regime (i.e., at $\rho\lesssim 1$ g~cm$^{-3}$
and $300\mbox{~K}\lesssim T \lesssim 2000$~K) and at high densities
($\rho\gg1$ g cm$^{-3}$), respectively, which, together with
$F_\mathrm{tran}$, provide the fit to the pressure as a function of
density through the thermodynamic relation 
\begin{equation}
P=-(\partial F/\partial V)_T,
\label{PbyF}
\end{equation}
and 
\begin{equation}
   w(\rho,T) =
   \frac{1}{1+(\rho/2.5\mbox{\,g\,cm}^{-3}+T/3509\,\mathrm{K})^4}
\end{equation}
is an interpolating function, which varies from
0 to 1 and ensures fitting the pressure as a function of density in the
entire $\rho-T$ domain considered.
In Eq.~(\ref{Fpl}), 
$F_\mathrm{e}(N_\mathrm{e},T,V)$ is the Helmholtz free energy of the ideal nonrelativistic
Fermi gas of $N_\mathrm{e}=n_\mathrm{e} V$ electrons at temperature $T$, and $Z_*$ is an
effective charge number, which is expressed as an analytical fitting
function so as to adjust the pressure derived through Eq.~(\ref{PbyF})
to the pressure data from the \textit{ab initio} calculations.
Explicitly,
\begin{equation}
   F_\mathrm{e}(N_\mathrm{e},T,V) = \mu_\mathrm{e} N_\mathrm{e} -
P_\mathrm{id}^\mathrm{(e)}\,V,
\end{equation}
where
\begin{equation}
   P_\mathrm{id}^\mathrm{(e)} =
    \frac{8}{3\sqrt\pi}\,\frac{\kB T}{ \lambda_\mathrm{e}^3}
     I_{3/2}(\mu_\mathrm{e}/\kB T)
\label{P_e}
\end{equation}
is the effective (ideal Fermi gas) electron pressure,
   $\lambda_\mathrm{e} = (2\pi\hbar^2/ m_\mathrm{e} \kB T)^{1/2}$ 
is the electron thermal wavelength,
\begin{equation}
   \mu_\mathrm{e} =\kB T X_{1/2}(n_\mathrm{e}\lambda_\mathrm{e}^3\sqrt{\pi}/4)
\label{mu_e}
\end{equation}
is the effective electron chemical potential,
and
\begin{equation}
  Z_* = \frac{10}{3}\,\Bigg(
       1 + \frac{2.35\,r_s}{1+0.09/(r_s\sqrt{\Gamma_\mathrm{e}})} 
        + \frac{5.9\,r_s^{3.78}}{(1+17/\Gamma_\mathrm{e})^{3/2}}
  \Bigg)^{-1}.
\label{Z*}
\end{equation}
In Eq.~(\ref{P_e}),
\begin{equation}
   I_\nu(X) \equiv \int_0^\infty
  \frac{ x^\nu\, \mathrm{d}x
    }{ \exp(x-X)+1 }
\end{equation}
is the standard Fermi integral, and $X_\nu(I)$ in Eq.~(\ref{mu_e}) is the
inverse function. For both $I_\nu(X)$ and $X_\nu(I)$ we use
the Pad\'e-type approximations of \cite{Antia93}.
In Eq.~(\ref{Z*}), 
$\Gamma_\mathrm{e}$ and $r_s$
are the usual electron Coulomb parameter and
density parameter, respectively:
$
   \Gamma_\mathrm{e} = e^2 / (a_\mathrm{e} \kB T)
$
in the CGS system, and
$
  r_s = a_\mathrm{e}/a_0 ,
$
where $a_\mathrm{e} = (\frac43\pi n_\mathrm{e})^{-1/3}$  is the
electron-sphere radius, and
$n_\mathrm{e}=10\,n_{\mathrm{H}_2\mathrm{O}}$ is the total number
density of all electrons (free and bound). In Eq.~(\ref{Flow}),
$a_\mathrm{vdW}$ and $b_\mathrm{vdW}$ are the van der Waals
constants (respectively, $5.524\times10^{12}$~erg cm$^3$ mol$^{-2}$ and 
$30.413$ cm$^3$ mol$^{-1}$, \citealp{handbook1997}). 
The first line in Eq.~(\ref{Flow}) reproduces the van der
Waals EOS, which is sufficiently accurate at $\rho\ll1$ g cm$^{-3}$, and
the second line adjusts the EOS at $\rho\sim1$ g~cm$^{-3}$. 

The fourth term in Eq.~(\ref{fit}) reads
\begin{eqnarray}
   F_T &=& N_\mathrm{at}  \left[ b_1 \tau \ln(1+\tau^{-2})
           + b_2 \tau \arctan\tau - b_3 \right]
\nonumber\\&&  \qquad
          + N_\mathrm{at} \kB T \ln\left[1+(0.019\tau)^{-5/2}\right],
\label{F_T}
\end{eqnarray}
where  $b_1=3\times10^{-13}$~erg, $b_2=1.35\times10^{-13}$~erg, $b_3=2.43\times10^{-13}$~erg,
and $\tau=T/T_\mathrm{crit}=T/647$~K. It is derived from the fitting
correction to the residual internal energy,
\begin{equation}
U_T = N_\mathrm{at} \frac{2b_1\tau-b_2\tau^2}{1+\tau^2}
    -b_3 N_\mathrm{at}
    +  \frac{2.5 N_\mathrm{at} \kB T}{1+(0.019\tau)^{5/2}}.
\end{equation}
This correction does not affect pressure but improves the fit to
the internal energy, through the thermodynamic relation
\begin{equation}
U=- T^2 \left. \frac{\partial}{  \partial T} \frac{F}{T} \, \right|_V.
\end{equation}
Note that we measure the total internal energy from its minimum at the ground state of the molecular phase,
so that $U>0$ at any
$\rho$ and $T$. This definition is the same as in \citet{Wagner_Pruss02}.
It differs from the definition adopted in
 Table~\ref{tab1} and Fig.~\ref{fig2} by the ground-state energy
value of 11.14 MJ/g.
It also differs by a constant of 77 kJ/g from the internal
energy given 
in \citet{French_etal09,French_Redmer15,Soubiran_Militzer15}.

The last term in Eq.~(\ref{fit}),
$-S_0T$ is an additional correction, which affects neither $P$ nor
$U$, but shifts the entropy 
\begin{equation}
S= - \, (\partial F/\partial T)_V
\end{equation}
by constant $S_0$. We find that the value
$S_0=9.8 \kB N_\mathrm{at}$ provides the best fit 
(within $\pm0.3\kB N_\mathrm{at}$)
to the results presented for $S$ by \citet{Soubiran_Militzer15} at
$\rho\approx(1-2.5)$ g~cm$^{-3}$ and $T=(1000-6000)$~K. 
However, entropy evaluation from our present fit should be used with
caution especially when crossing the boundaries of different phases,
where one can expect a discontinuity. This may lead to a value of $S_0$ that differs from one phase to the other.

The present analytical fit describes the EOS of liquid water at $\rho \lesssim
1$ g~cm$^{-3}$ and $T\lesssim 2000$~K, as well as plasma at $\mbox{1
g~cm$^{-3}$} \lesssim\rho\lesssim10^2$ g~cm$^{-3}$ and $10^3\mbox{ K}\lesssim
T\lesssim \times10^5$~K. While not including the super-ionic phase as a
different phase, the single-phase approximation used here provides a
satisfactory description of the thermodynamical properties in this super-ionic
regime. It has, however, a limited applicability for the ice VII and ice X
phases that occurs at $T\lesssim2000$~K in the range
$(0.02-0.5)\mbox{~Mbar}\lesssim P \lesssim 3$~Mbar \citep{ice1999}.
We also point out that quantum effects for the ions, which could be
  relevant at the highest densities and for low temperatures, are not
  included in the current parametrization.
To build a fully multi-phase EOS for water, one can supplement our fit by the
parametrizations constructed specifically for the ice and super-ionic phases
\citep{French_Redmer15,French_Desjarlais_Redmer16}. This will be the topic of
further work. Finally, we also point out that our analytical fit is less
reliable in the domain of thermal ionization and dissociation of molecular
water, where $\rho\ll1$ g~cm$^{-3}$ and $T\gg10^3$~K. This regime is indeed
poorly constrained by either the \textit{ab initio} simulations or the IAPWS
parametrization.

%%%%%%%%%%%%%%%%%%%%%%%%%%%%%%%%%%%%%%%%%%%%%%
\subsection{Validation of the analytical fit}
 
\begin{figure}
   \centering
   \includegraphics[width=\columnwidth]{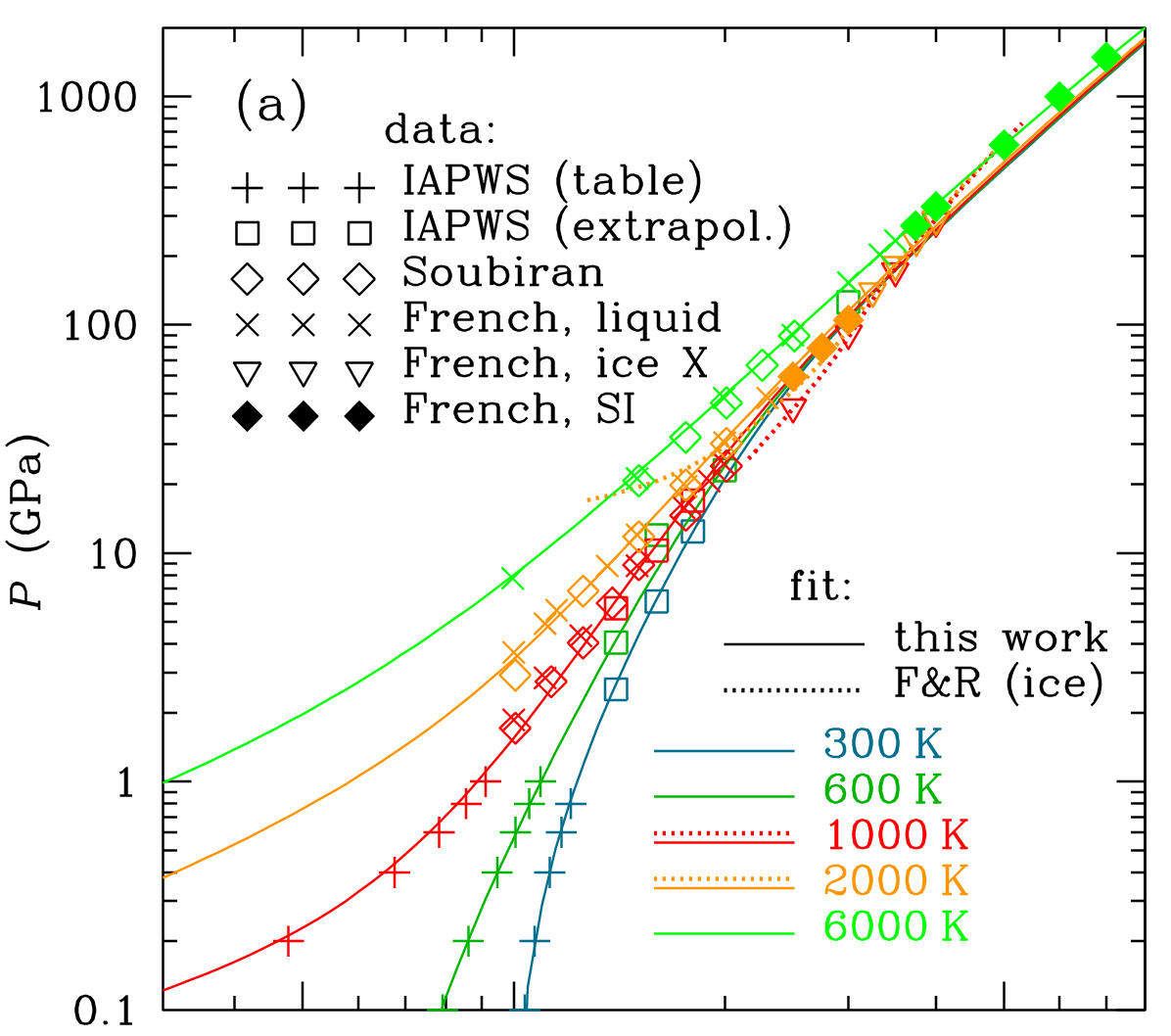}
   \includegraphics[width=\columnwidth]{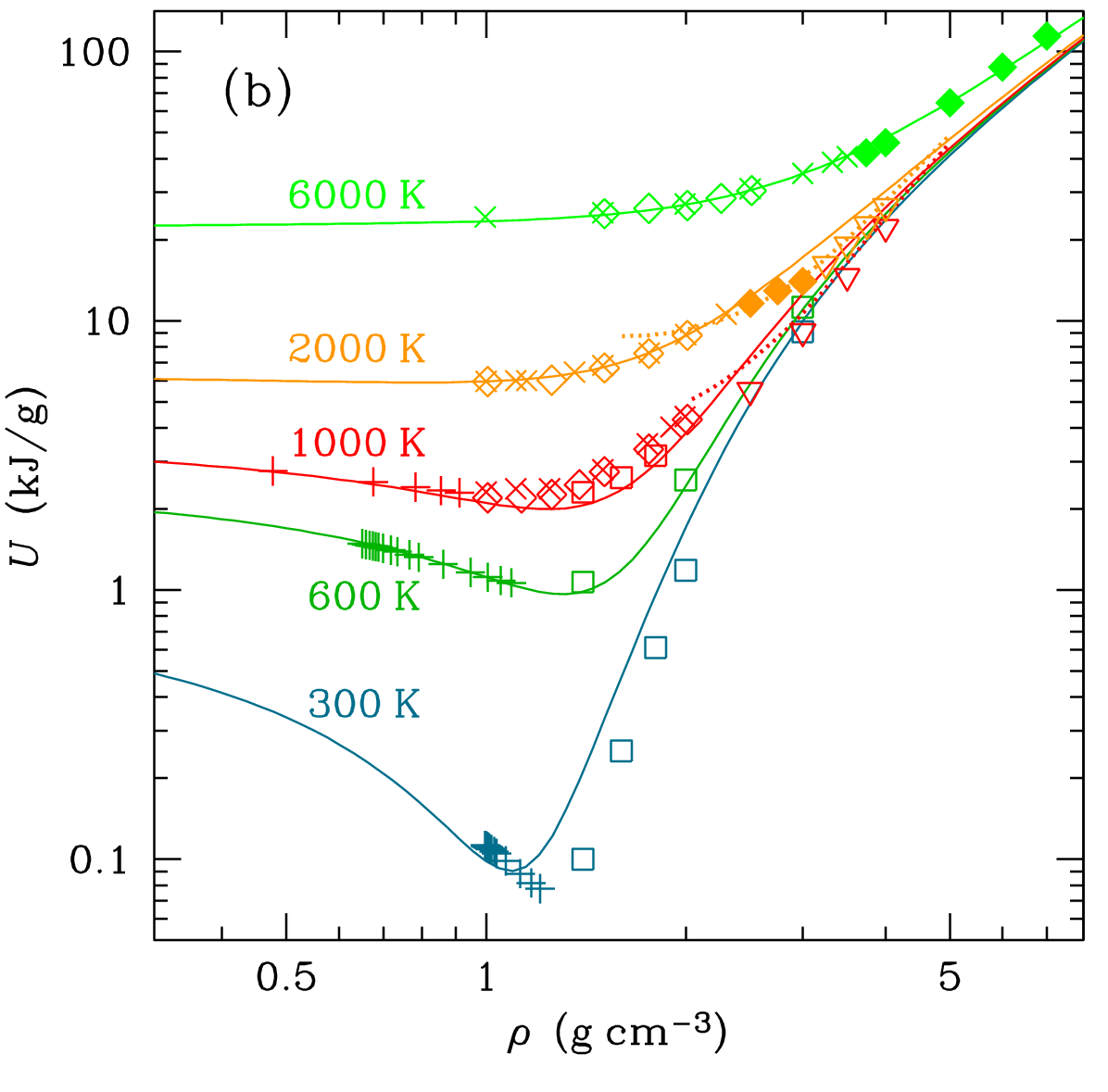}
   \caption{Comparison between the 
   input data  and the analytical fitted isotherms for the
   pressure $P$ (a) and for the internal energy $U$ (b)
   at relatively low densities. Symbols show the data:
   the IAPWS \citep{Wagner_Pruss02} published table for $P<1$~GPa
    (straight crosses)
   and extension to $P>1$~GPa according to the IAPWS
   free-energy model (squares); results of 
   \textit{ab initio} calculations by \citet{Soubiran_Militzer15} (empty
   diamonds) and
   by \citet{French_etal09} (oblique crosses for liquid, inverted
   triangles for ice X, filled diamonds for the superionic [SI] phase). 
   Solid lines represent the present fit; dotted lines
   represent the fit of \citet{French_Redmer15} for ice X.}
 \label{fig3}%
  \end{figure}

\begin{figure}
   \centering
   \includegraphics[width=\columnwidth]{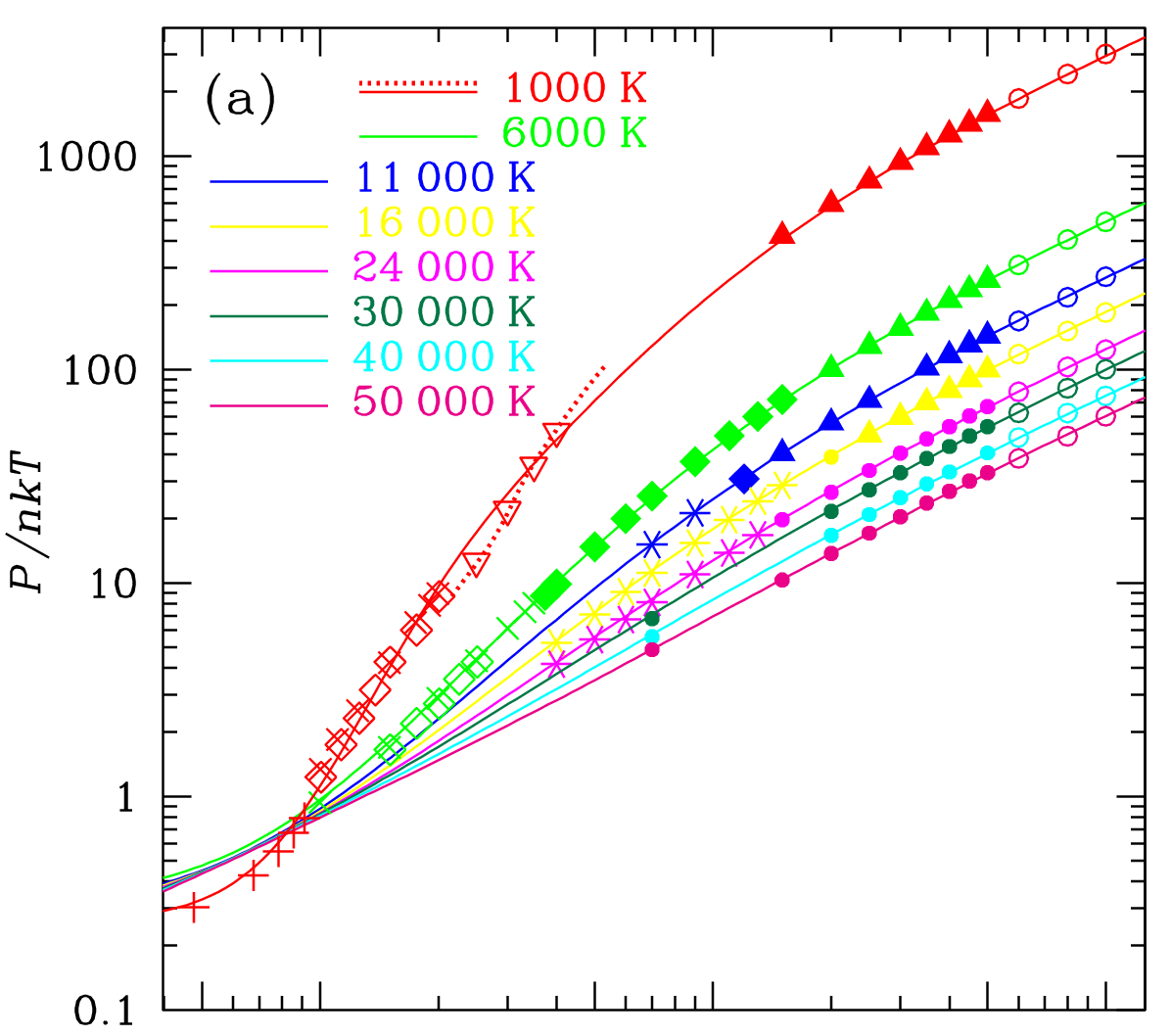}
   \includegraphics[width=\columnwidth]{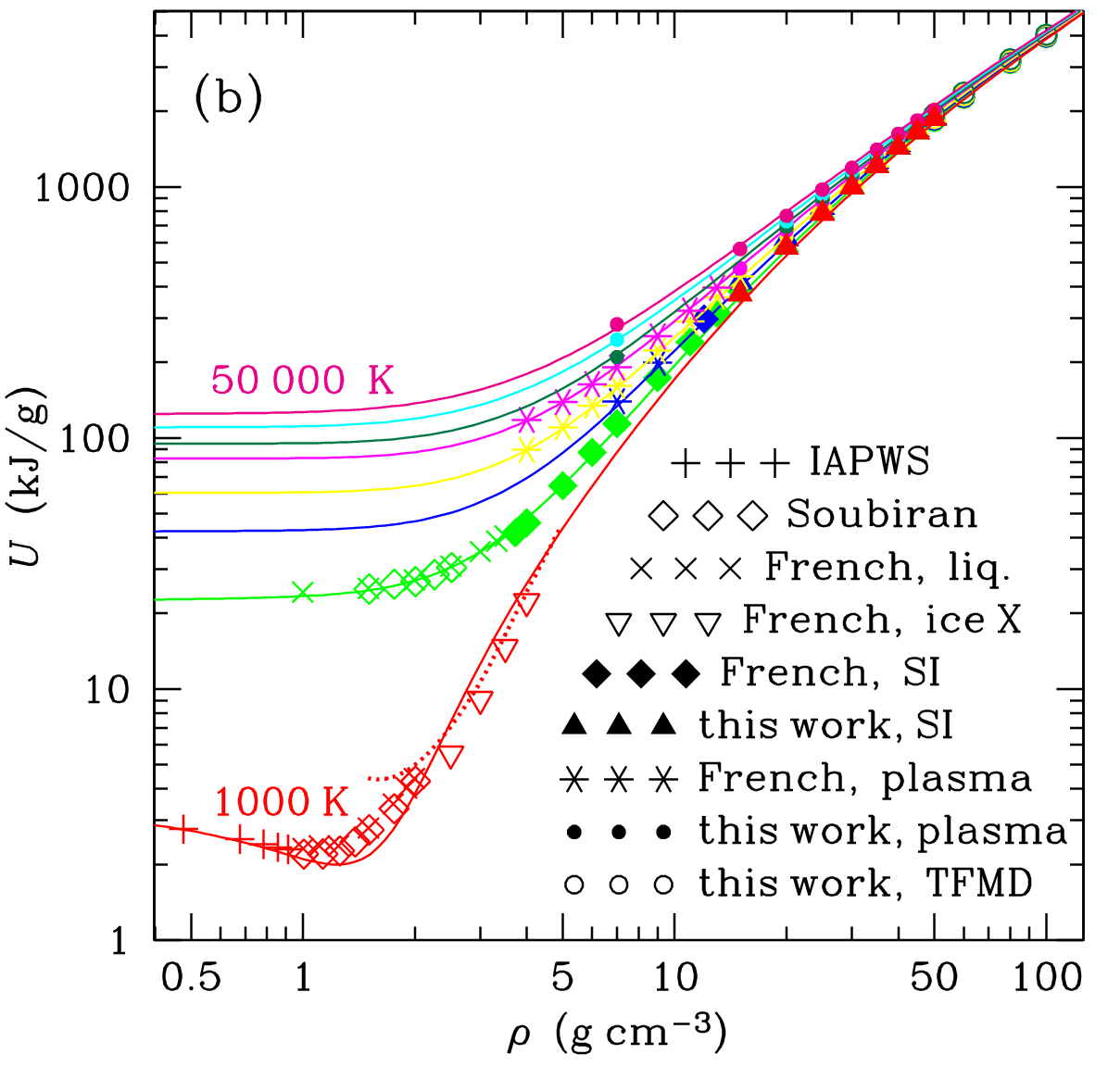}
   \caption{Comparison of the data with
   the analytical fitted isotherms
   at high densities for the
   pressure normalized to the ideal atomic gas value, $P/n_\mathrm{at} \kB T$ (a),
    and for the internal energy $U$ (b). Symbols show the data:
   the IAPWS table (\citealp{Wagner_Pruss02}, straight crosses);
   results of \textit{ab initio} calculations
   by \citet{Soubiran_Militzer15} (empty
   diamonds); \textit{ab initio} calculations
   by \citet{French_etal09} (oblique crosses for liquid, inverted
   triangles for ice X, filled diamonds for the superionic [SI] phase,
   asterisks for the plasma phase); and the results of
   our present \textit{ab initio} calculations (filled triangles for the
   SI phase and filled dots for the plasma phase), supplemented with our
   TFMD calculations at $\rho>50$ g~cm$^{-3}$ (empty circles).
    Solid lines represent the present fit; the dotted line
   represents the fit of \citet{French_Redmer15} for ice X at $T=1000$~K.}
\label{fig4}%
  \end{figure}

We first verify the ability of the analytical fit at reproducing both the
results of theoretical calculations and the IAPWS free energy model. In
Figs.~\ref{fig3} and \ref{fig4}, we compare the behavior of pressure and
internal energy obtained with the input data at respectively
low and high densities and temperatures.

  As we are primarily interested
in planetary interiors, the accurate description of the liquid-vapor
transition below the critical point located at $T_\mathrm{crit}=647$~K and
$P_\mathrm{crit}=22.064$~MPa is beyond the scope of this study.
Furthermore, these conditions are tied to an accurate modeling of the
atmosphere of the planet that do not directly involve an EOS as the one
developed here. Figures \ref{fig3}a-b suggest that without atmospheric
treatment, interior models should consider the liquid state for surface
conditions when the surface temperature is below the critical point. We
will expand further on this point in the following sections.

At the lowest densities, we see
in figure \ref{fig3}a that the pressure turns negative along the 300~K and
600~K isotherms for densities below 1 g~cm$^{-3}$. Figure~\ref{fig3}b
indicates that this translates into a minimum for the internal energy.
This corresponds to the crossing of the liquid-vapor phase boundary and a
region where the pressures are formally negative, which in fact corresponds
to phase coexistence. For instance, for the 300~K isotherm 
the analytical fit gives a
region of negative pressures expanding from 0.9 g~cm$^{-3}$ to 0.1
g~cm$^{-3}$. Note that the IAPWS data \citep{Wagner_Pruss02} indicate a
wider phase-transition region, with the low density $\sim10^{-3}$
g~cm$^{-3}$ on this isotherm.
The agreement improves for
the 600~K isotherm. We see that the analytical fit reproduces the
overall behavior of the pressure across the liquid-vapor boundary.
However, it extends this boundary to a higher temperature, giving the
critical point at $683$~K and 0.331 g cm$^{-3}$ (to be compared with the
experimental values of 647~K and 0.322 g cm$^{-3}$).

Figures \ref{fig3}a-b also indicate that the agreement with the \textit{ab
initio} data at higher temperatures is satisfactory up to 6000~K. We note
that the \textit{ab initio} results and the free energy model predictions
are not in perfect agreement at 1000~K. As the \textit{ab initio} method
becomes less reliable as density decreases, mainly due to the deficiency
of density functional theory in under-dense regime, the \textit{ab initio}
results fail to match the IAPWS formulation at low density. Our analytical
fit eliminates this mismatch by interpolation between the low-density
IAPWS and high-density \textit{ab initio} data.

Figures \ref{fig4}a-b show the data and fitted isotherms for the pressure
and internal energy across the entire density range at higher
temperatures, $1000\mbox{~K}\leq T\leq50\,000$~K. Figure \ref{fig4}a
displays the pressure normalized to the atomic ideal gas contribution
$n_\mathrm{at} \kB T$. As noted above, the fit does not perfectly
reproduce the \textit{ab initio} data for the ice X phase along the 1000~K
isotherm at 2.5 g~cm$^{-3} \leq\rho \leq4$ g~cm$^{-3}$. However, it
satisfactorily reproduces the thermodynamical properties, despite the fact that the system
crosses a number of various phases in the  $\rho-T$ domain displayed in
the figures. 

We also point out that the data represented in Figs.~\ref{fig3} and
\ref{fig4} by empty symbols (the ice phase, the data by
\citealt{Soubiran_Militzer15}, and the Thomas-Fermi results) have not been explicitly
included to construct the analytical fit. We see that the data by
\citet{Soubiran_Militzer15} is in good agreement with the prediction of the analytical fit. 
Furthermore, the good agreement found with the Thomas-Fermi results up to 100 g~cm$^{-3}$ 
shows the validity of the analytical fit up to this high density.
\begin{figure}
   \centering
   \includegraphics[width=\columnwidth]{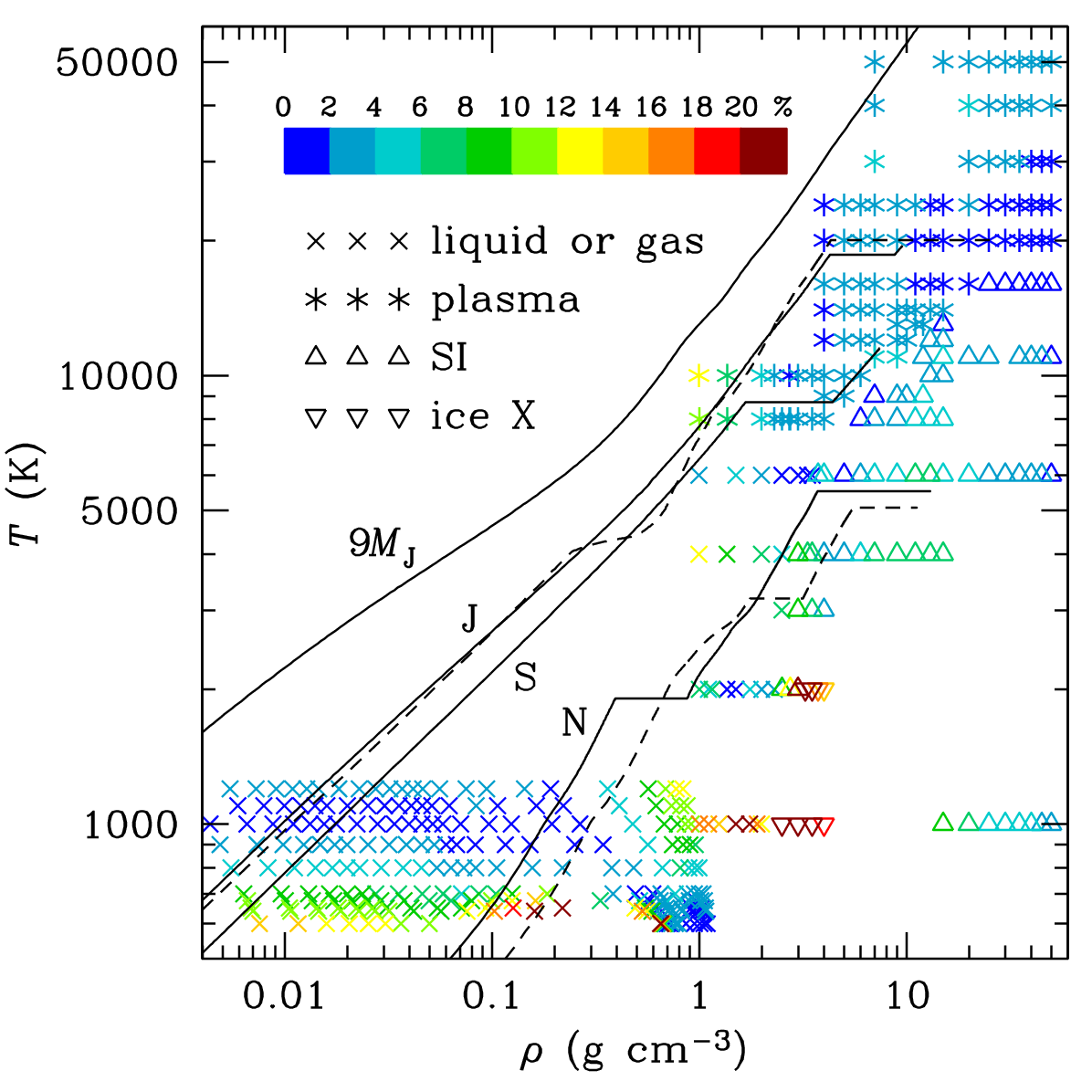}
   \caption{Points in the $\rho-T$ plane where the input data have
been used to construct the analytical fit. Different symbols correspond to
different phase states: crosses for liquid, asterisks for plasma, upright
triangles for superionic state, and reverted triangles for ice X. The colors
of the symbols represents the accuracy of the fit (for \emph{both} $P$ and $U$,
i.e., the maximum of the two residuals) according to the palette above the
legend. 
The lines show isentropes of Jupiter (J), according to the models of
\citet{Leconte_Chabrier12} and \citet{Nettelmann_etal12} (the solid and dashed
lines, respectively), Saturn (S), according to \citet{Leconte_Chabrier12},
Neptune (N) according to  two models of \citet{Nettelmann_etal13} (solid and
dashed lines), and a planet with $M=9\,M_J$ 
(see \citealt{Baraffe_Chabrier_Barman08,Baraffe_Chabrier_Barman10}).
}
\label{fig:fitdomain}
\end{figure}

Fig.~\ref{fig:fitdomain} summarizes the applicability of the current analytical
EOS  for planetary modeling. We display the accuracy of the analytical fit as a
color code corresponding to the residual difference between the predicted and
input $P$ and $U$ values. On the same figure,  we also show representative
interior profiles for Jupiter, Saturn, and Neptune that display a significant
amount of water in their interior. This shows that the analytical fit is
accurate for modeling these objects. For comparison, the profile of a  9$M_J$ planet is also shown. As pointed out before, the single-phase
approximation used for this analytical fit ignores the discontinuities due to the
phase changes between the different phase states. For the phase transition between the liquid state and ice X,
the discrepancies increase to tens percent.  Otherwise we see that the fit
remains a reasonable approximation for both the pressure or internal energy
throughout the relevant thermodynamical domain.

%%%%%%%%%%%%%%%%%%%%%%%%%%%%%%%%%%%%%%%%%%%%%%%%%%%%%%%%%%%%%
\section{Comparison with experimental data and previous EOSs}

\begin{figure}
   \centering
   \includegraphics[width=\columnwidth]{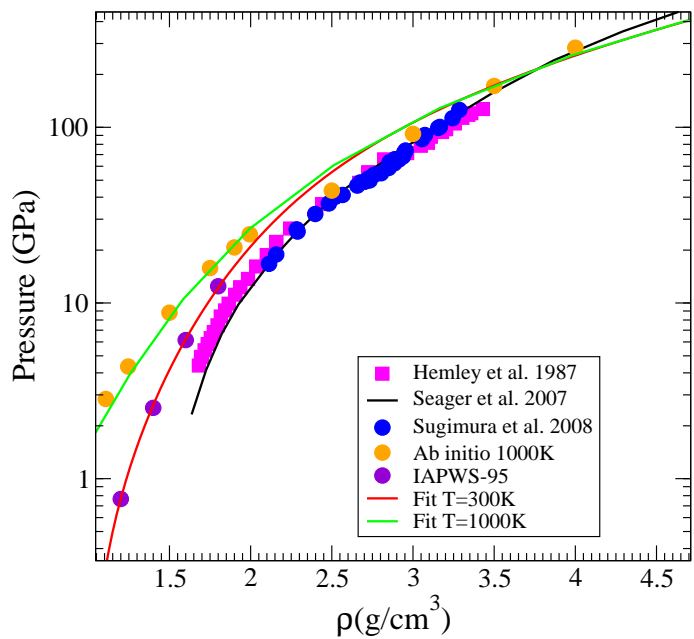}
   \caption{Comparison of the analytical fit predictions with high-pressure data at $T=300$~K.}
              \label{fig:static}%
\end{figure}

We now turn to compare the predictions of the analytical fit
developed in Sect.~\ref{sect:fit} with existing experimental data from both
static and dynamical experiments. We compare these predictions with
EOSs commonly used in planetary interior models. In Fig.~\ref{fig:static},
we compare the predictions of the analytical fit developed with
the static high pressure data obtained using diamond anvil cells
\citep{Hemley_etal87,Sugimura_etal08}. We also show in
Fig.~\ref{fig:static} the 1000~K isotherm and its corresponding \textit{ab
initio} data to illustrate further the approximation made by not
accounting explicitly for the solid ice phases and extending the liquid
throughout the solid phases. We see that the analytical fit thus misses the
jump between the liquid and ice X phases at $\rho\sim2.5$ g~cm$^{-3}$, as
we have already seen in Fig.~\ref{fig4}.

 The $T=300$~K isotherm behaves similarly. We see that both the
IAPWS free energy formulation and our  analytical fit, being continued
from the low-density region at $T=300$~K, overestimate the pressure
(underestimate the density) in the ice phases at higher densities. At
$P=10$~GPa, the resulting density  is about 7\% lower than in the ice
phases. This should be compared with the predictions of 
\citet{Seager_etal07} who used $T=0$~K DFT calculations for the various
ice phases to construct their EOS. By neglecting the effect of
temperature  and the phonon contribution, they overestimate the density of
dense water by about 3\% at $10$~GPa.  We also note that this difference
with the experimental data tends to decrease as the pressure increases for
both EOSs. Therefore, the most significant difference resulting from
neglecting the ice phases that we can anticipate for interior structure
calculations could be for the pressure profile of the outer
layer of a planet, if it had $T\sim10^3$~K at $\rho\sim3$ g cm$^{-3}$.
From Fig.~\ref{fig:fitdomain} we see that it is not the case for the giant
planets, for which the temperature is much higher and thus the temperature
profile passes well above this phase jump.

We also point out some differences between the two \textit{ab
initio} calculations beyond 4 g cm$^{-3}$. We attribute this difference to the
use of different functionals  in the Thomas-Fermi approximation. We will
investigate further its impact on the interior structure models in the
following sections.

\begin{figure}
 \centering
 \includegraphics[width=9cm]{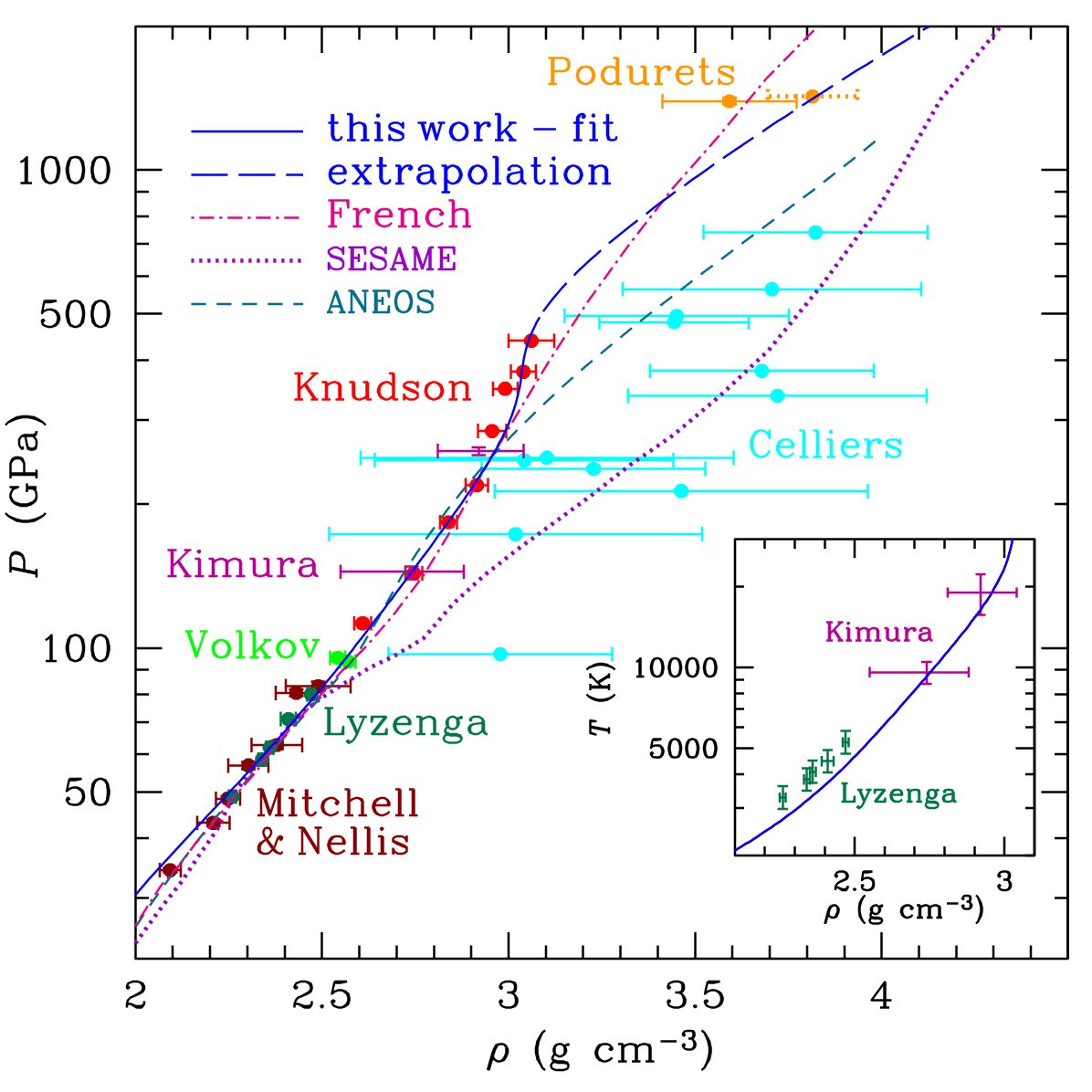}
 \caption{The principal Hugoniot line in the $\rho-P$ plane calculated
using the present fit (solid line for $T<50\,000$~K, continued by
long-dashed line for $T>50\,000$~K) compared with experimental data (dots
with error bars) of \citet{Podurets_etal72} (as reanalyzed by
\citealt{Knudson_etal12}; the original result of
\citeauthor{Podurets_etal72} is also shown with dotted errorbars),
\citet{Volkov_etal80,Mitchell_Nellis82,Lyzenga_etal82,Celliers_etal04,Knudson_etal12},
and \citet{Kimura_etal15}. For comparison, the principal Hugoniot lines
predicted by the SESAME \citep{sesame} and ANEOS \citep{aneos} models are
shown by the dotted line and short-dashed line, respectively. The inset
shows the principal Hugoniot line in the $\rho - T$ plane calculated from
the fit and the experimental data points from \citet{Lyzenga_etal82} and
\citet{Kimura_etal15}.}
\label{fig:shock}
  \end{figure}

Figure \ref{fig:shock} shows a comparison between the predictions of our
analytical fit deduced from \textit{ab initio} simulations and the
measured experimental data along the principal shock Hugoniot line. For a
given initial state, the Rankine-Hugoniot relation determines the final
states allowed by conservation of energy and momentum during a shock. It
is directly related to the EOS and reads
\begin{equation}
   U-U_0=\frac{P+P_0}{2}(V_0-V),
\label{Hugoniot}
\end{equation}
where subscript 0 indicates the initial state. The \emph{principal
Hugoniot} corresponds to a single shock, obtained with initial state at
rest at normal conditions. Figure \ref{fig:shock} shows that shock  experiments
probe a range of pressures almost an order of magnitude higher than when
using diamond anvil cells (Fig.~\ref{fig:static}). This also corresponds to a
significant increase in temperature. The temperature reaches about
5\,000~K around 100~GPa for a shock,
while it remains near 300~K in diamond anvil cell experiments. With the
increase of the shock pressure beyond $500$~GPa, the temperature exceeds
$50\,000$~K.  Since the input \textit{ab initio} data that underlie our fit
have been obtained for $T\leq50\,000$~K, the accuracy of the fit at higher
temperatures is not guaranteed (the corresponding part of the Hugoniot
line is drawn by long dashes in Fig.~\ref{fig:shock}).

Figure \ref{fig:shock} shows a good agreement at low pressures with early
data obtained using explosion \citep{Volkov_etal80} and gas-gun techniques
\citep{Mitchell_Nellis82,Lyzenga_etal82}. The unique
measurement of the water EOS at $P>1000$~GPa, published long ago by
\citet{Podurets_etal72}, is satisfactorily described by the
above-mentioned continuation of our fit beyond the range where it has been
constrained by the data. The
first laser shock data obtained in the 100 to 1000~GPa range by
\citet{Celliers_etal04} are much softer than the analytical fit. This
experimental data set is in good agreement with SESAME 7150 predictions.
In contrast, we see that the analytical fit agrees very nicely with the
more recent experimental data of \citet{Knudson_etal12} obtained using the
Z-pinch techniques, as well as with the latest results of
\citet{Kimura_etal15}. This confirms that earlier laser shock experiments
likely suffer from systematic errors which could be caused by the standard
used in the impedance matching method \citep{Knudson_Desjarlais09}. This
experimental set will thus not be considered to validate the behavior of
water at high pressures and temperatures. This also highlights that the
SESAME 7150 predictions, that are in good agreement with the data of
\citet{Celliers_etal04}, should be ruled out for planetary modeling. The
SESAME 7150 model is not considered a reliable EOS nowadays, but it is
still in use for planetary modeling  \citep{Miguel_Guillot_Fayon16}. This
is also the case for the ANEOS model \citep{aneos}.
Figure~\ref{fig:shock} shows that ANEOS predictions depart from the data
of \citet{Knudson_etal12} at $\rho>2.5$ g~cm$^{-3}$ significantly outside
the experimental errorbars.

\begin{figure}
 \centering
 \includegraphics[width=9cm]{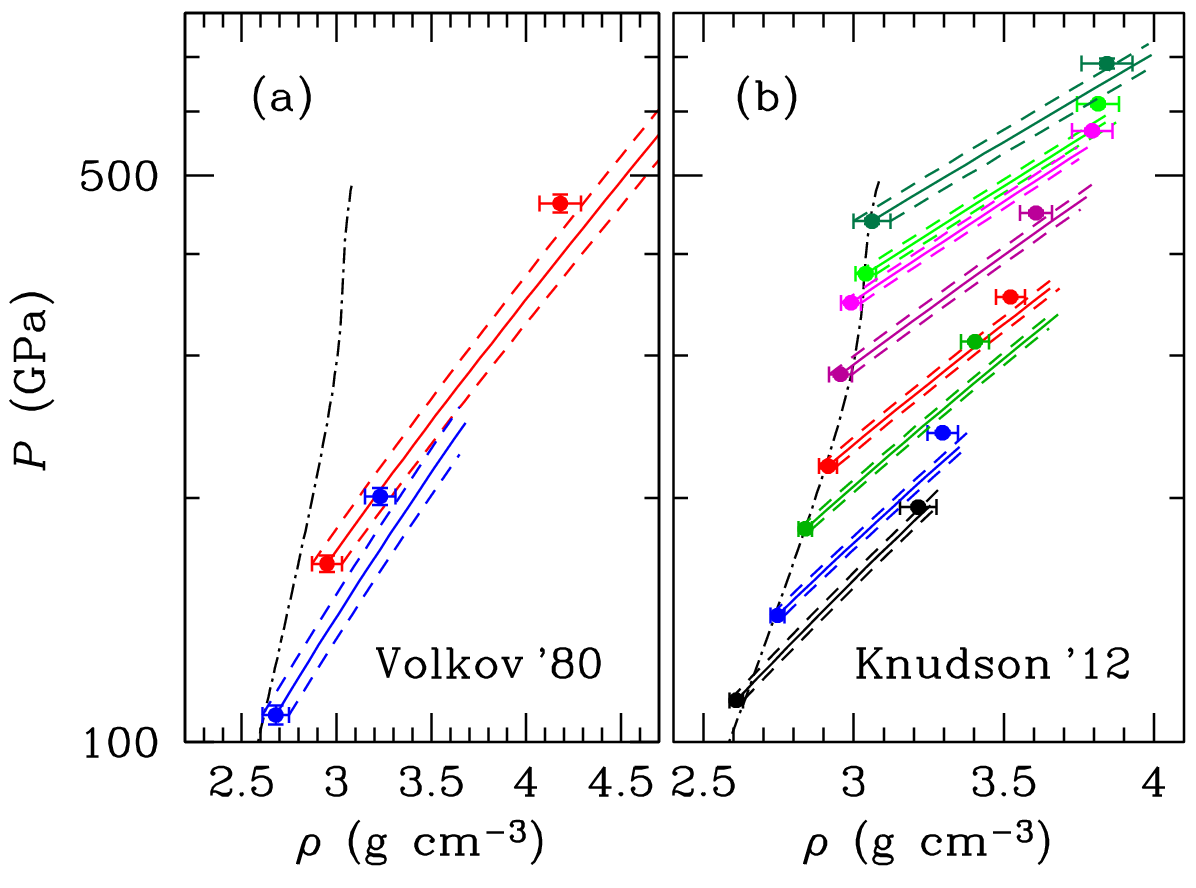}
 \caption{Comparison of experimental data for reshocked water (points with
errorbars) with corresponding Hugoniot lines calculated from the analytical fit
(solid lines). The principal Hugoniot is shown by the dot-dashed line.
Dashed lines show the theoretical regions defined by taking into account $1\sigma$
experimental uncertainties for the initial (primary-shock)
states. Panels (a) and (b) show the comparison for the data of 
\citet{Volkov_etal80} and of \citet{Knudson_etal12}, respectively.}
\label{fig:reshock}
  \end{figure}

Figure \ref{fig:reshock} shows a comparison between the
predictions of our analytical fit and the EOS measurements obtained using
double-shock experiments \citep{Volkov_etal80,Knudson_etal12}, where the
initial shock wave is reflected from a surface of a standard material
(aluminum or quartz). In this case, the initial state in
Eq.~(\ref{Hugoniot}) lies on the principal Hugoniot line and represents the initial condition for the
secondary Hugoniot. Since the latter initial condition
is measured with some uncertainties, the position of the 
secondary Hugoniot is not firmly
defined. We therefore show the 1$\sigma$ limits for each secondary
Hugoniot that arise from these uncertainties in the initial state.

\begin{figure}
 \centering
 \includegraphics[width=9cm]{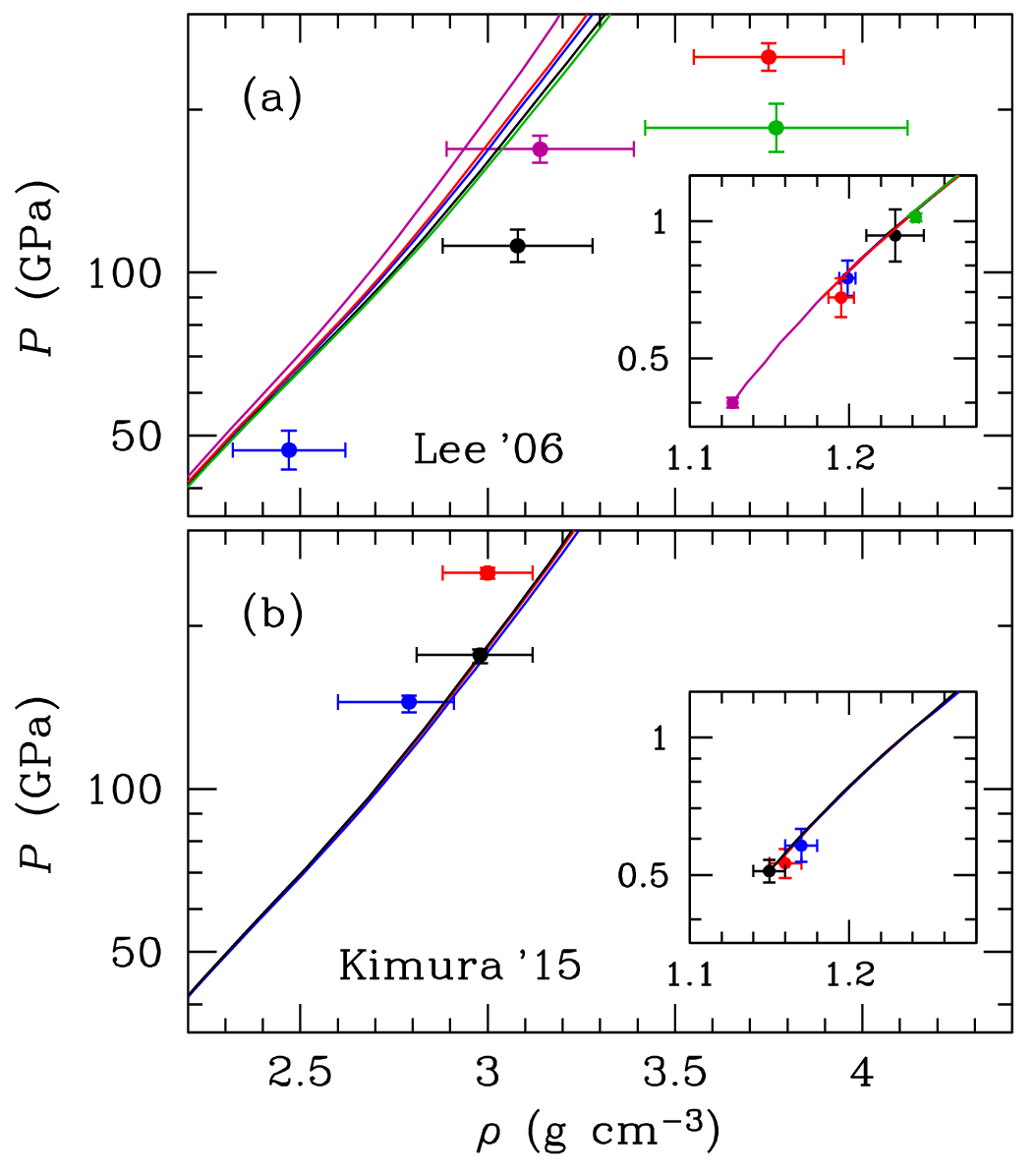}
 \caption{Comparison between the calculations based on our analytical fit and
the experimental shock data for initially precompressed water. (a) Points
measured by \citet{Lee_etal06}; (b) points obtained by \citet{Knudson_etal12}.
The inset in each panel shows the initial pressures and densities measured.}
\label{fig:precompr}
  \end{figure}

Figure \ref{fig:precompr} shows a comparison between the predictions of
our analytical fit with two EOS measurements in experiments using laser-induced shock
 in statically compressed water. We find a better agreement with the latest dataset of \citet{Kimura_etal15}
 compared to the ones of \citet{Lee_etal06}. We show these for completeness as the scatter in the experimental data
 cannot further constrain the validity of the analytical fit.

Overall, we find that our present analytical fit provides a satisfactory
description of both the static and dynamical experimental results
available to date for dense water. We now turn to illustrate the
applications of the EOS developed for the interior
structure of different classes of planets. 

%%%%%%%%%%%%%%%%%%%%%%%%%%%%%%%%%%%%%%%%%%%%%
\section{Implication for planetary interiors}

The internal structure calculations are performed by solving the standard
hydrostatic, mass, and energy conservation equations
\citep[e.g.,][]{Schwarzschild58,kippenhahn2012stellar} 
\begin{eqnarray}
  \frac{\partial P}{\partial r} &=& -\rho g,
\label{eqhyd1}
\\
\frac{\partial T}{\partial r} &=&
 \frac{\partial P}{\partial r}\frac{T}{P}\nabla_T,
\\
\frac{\partial m}{\partial r} &=& 4\pi r^2\rho,
\label{eqhyd3}
\end{eqnarray}
where $P$ is the pressure, $\rho$ the density, $g=Gm/r^2$ the gravity, 
$G$ is the gravitational constant, $m$ is the mass enclosed within a
sphere of radius $r$, and  
$\nabla_T=\mathrm{d}\ln T / \mathrm{d}\ln P$ depends on the mechanism of
energy transport.  If the medium is stable against convection, then
\begin{equation}
    \nabla_T = \frac{3}{16}\,\frac{PK}{g}\,\frac{T_\mathrm{eff}^4}{T^4},
\end{equation}
where $T_\mathrm{eff}$ is the effective surface temperature 
and $K$ is the effective opacity.  If the transport of
energy is dominated by convection, then in the simplest (Schwarzschild)
approximation $\nabla_T=\nabla_\mathrm{ad}$, where
\begin{equation}
  \nabla_\mathrm{ad}=\left.\frac{\partial \ln T}{\partial \ln P}\right|_S
  \label{adia}
\end{equation}
is the adiabatic gradient.

The interior structure of a planet of a given mass, $M$, is obtained by integrating inward the set of equations (\ref{eqhyd1})--(\ref{eqhyd3}),
starting from a
boundary condition defined by a fixed surface temperature and pressure. This is formally given by either in-situ measurements for planets of the
solar system or full atmospheric calculations for exoplanets. As we are primarily interested in testing our EOS for dense water, we will devise
strategies to overcome this difficulty for each type of planets that we will consider.

\subsection{Water planets}

\begin{figure}
   \centering
   \includegraphics[width=\columnwidth]{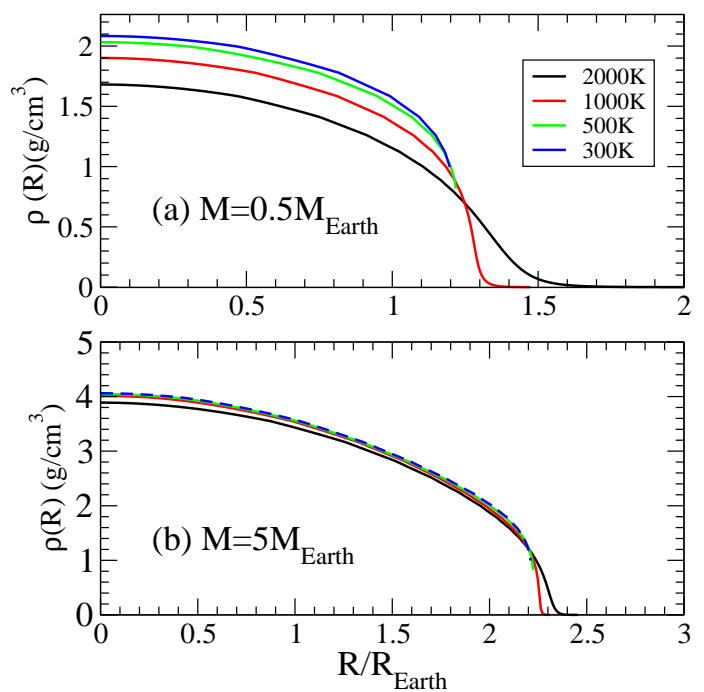}
   \caption{ Isothermal density profiles of pure water planets for
   varying surface temperature. The values of the surface temperature,
   $T_\mathrm{surf}$,  are indicated in the figure: (a)   for a planet of
   mass $M=0.5M_\mathrm{Earth}$, (b) same for a planet of mass
   $M=5M_\mathrm{Earth}$ as indicated  in the figure. $R_\mathrm{Earth}$
   stands for the Earth radius.}
              \label{fig7}%
 \end{figure}

In order to test the accuracy of our EOS, we first turn to the calculation
of the inner structure and mass-radius relationship of a planet entirely
made of water. This is a purely academic exercise disconnected from the
outcomes of planetary formation but the resulting mass-radius relationship
remains a well used benchmark to classify exoplanets
\citep{Seager_etal07}. It also allows us to decipher the influence of
temperature on both the inner structure and mass-radius relationship.
Figure \ref{fig7} shows the effect of temperature in isothermal interior
structure models for planet of respectively 0.5 and 5 $M_\mathrm{Earth}$.
In this situation, the temperature throughout the planet is constant and
equal to the surface temperature,  $T_\mathrm{surf}$. The outer boundary
conditions are chosen as $1$~bar for temperature above the critical point
and at the liquid-vapor boundary below. This corresponds to different
surface densities and close to 1 g~cm$^{-3}$ for the two models using a
surface temperature below the critical value.

Figure \ref{fig7} shows that within this model, the effect of temperature
is twofold. As the EOS developed fully accounts for temperature
throughout the density range covered by the conditions existing in
planetary interiors, it affects the extent of the outer edge boundary
as well as the compressibility deeper within the planet. Figure~\ref{fig7}
shows that both these effects are important for low mass planets where we 
notice a radius increasing by a factor of two when the surface temperature
varies from 300~K to 2000~K. This increase in radius comes from a
significant expansion of the low density outer edge beyond
$1.5R_\mathrm{Earth}$ that follows the rapid expansion of the
supercritical liquid at high temperatures and low surface gravity. It also
comes from the significant temperature dependence of the water EOS for
densities below 2 g~cm$^{-3}$ previously pointed out in Fig.~\ref{fig:static}. For
a planet of $0.5M_\mathrm{Earth}$, the maximum pressure reached at the
center of the planet is 0.18~Mbar for a temperature of
2000~K and 0.27~Mbar for a temperature of 300~K.  

   Figure \ref{fig7} shows that the effect of temperature decreases significantly when the planet size increases. This comes from both a smaller expansion
   of the outer edge that comes from a larger surface gravity and a reduced
   temperature dependence of the EOS as the density increases. Figure~\ref{fig7}
   shows that the density profile of the planet is dominated by densities
between 2 g~cm$^{-3}$ and 4 g~cm$^{-3}$ for a planet of $5 M_\mathrm{Earth}$. Figure \ref{fig:static}
shows
   that the temperature dependence of the EOS is almost negligible in this density range. The effect of temperature is thus reduced to a small expansion of the
   outer edge that leads to an increase of 10\% of the radius. This suggests that the effect of temperature becomes negligible as the mass of the planet
   increases.    

\begin{figure}
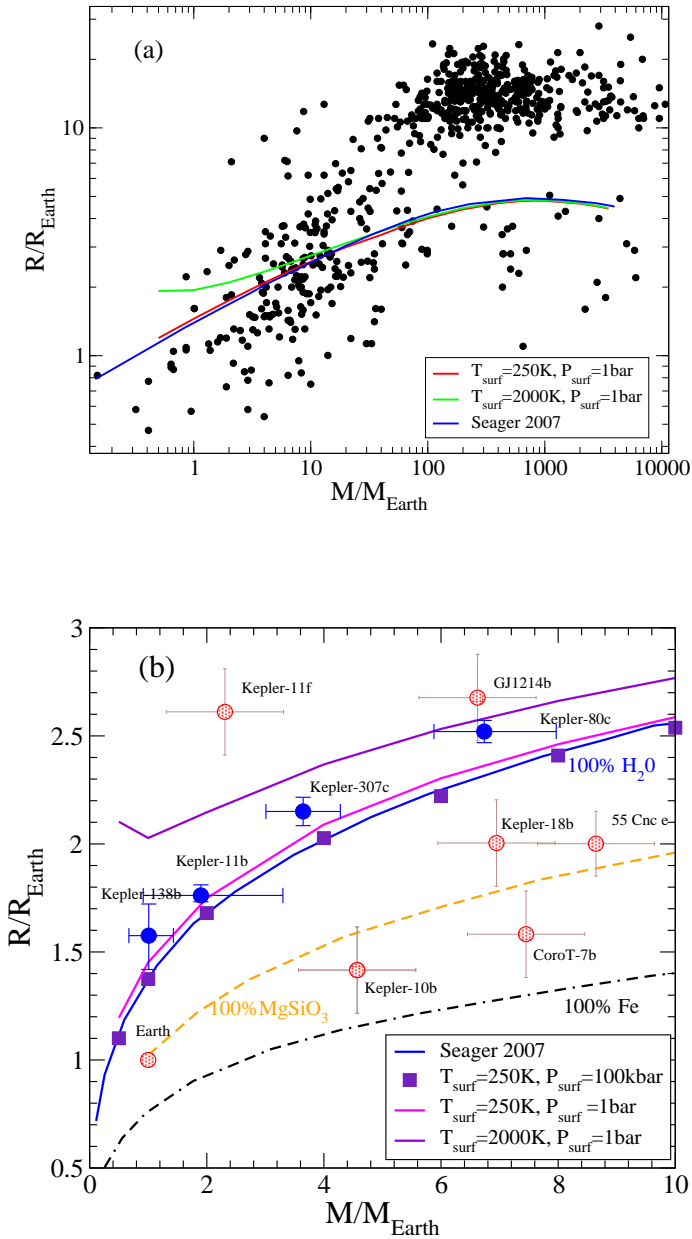

   \centering
   \includegraphics[width=\columnwidth,bb= 3 -50 528 422]{fig8-a.eps}
   \includegraphics[width=\columnwidth]{fig8-b.eps}
   \caption{Mass-radius relationship for pure water planets in an
   isothermal model. The surface temperature and pressure
    are indicated in the figure and compared to the result of \citet{Seager_etal07}. (a) over the entire range of planets 
    currently detected. (b) Same as (a) for planets with mass less than
$10M_\mathrm{Earth}$.  \textit{Isothermal models} for pure silicates (MgSiO$_3$), iron
(Fe) 
    \citep{Bouchet_etal13,Mazevet_etal15} }
              \label{fig8}%
\end{figure}

Figures \ref{fig8}a and \ref{fig8}b give the mass-radius relationship
obtained with the current EOS and calculated for the complete range of
planets detected to date, and a zoom for planets below
10\,$M_\mathrm{Earth}$ \citep{exoplanet.eu}. As anticipated above, we see
that the temperature dependence decreases as the mass of the planet
increases. Figure~\ref{fig8}b shows that the temperature effect is rather
important for planets smaller than 10$M_\mathrm{Earth}$. We also see in
figure \ref{fig8}a that the temperature dependence for pure water planets
can be neglected for planets larger than $15M_\mathrm{Earth}$ when
considering isothermal temperature profile. We also find that the
mass-radius relationship obtained with the EOS developed in the current
work is consistent with the calculations of \citet{Seager_etal07} for the
entire range of planets detected. Figures~\ref{fig8}a and \ref{fig8}b show
that the radius obtained with the EOS developed here and for an isothermal
model with surface temperature of $250$~K is 3\% higher than the results
of \citeauthor{Seager_etal07} for planets less than $10M_\mathrm{Earth}$
and 2\% lower above this planetary mass. This latter result comes from the
difference in the behavior of the two EOS at high densities already
mentioned in the previous section. For low mass planets, where inspection
of Fig.~\ref{fig:shock} indicates the largest difference between the EOS
developed here with both the experimental data and the zero temperature
EOS of \citeauthor{Seager_etal07}, we also find a rather satisfactory
agreement. This indicates that neglecting the ice phases has a minimal
impact on the resulting mass-radius relationship even for low mass
planets. For benchmarking purposes, we also show in Fig.~\ref{fig8}b that
the zero temperature results of \citeauthor{Seager_etal07} can be
recovered with the current EOS by considering a surface pressure of 10~GPa
(0.1~Mbar). 

Finally, we notice in Fig.~\ref{fig8}b that several planets detected fall within the temperature dependent range of the isothermal pure water model.
This suggests that temperature-dependent effect on the mass-radius relationship needs to be included when considering the interior structure to identify
the nature of these objects. We will pursue further this suggestion using more realistic interior structure models and by considering wet super-Earths and ocean
planets. 

\subsection{Wet super-Earth planets or ocean planets}

 Wet super-Earth or ocean planets are objects that have no
equivalent in the solar system. They have been introduced to interpret the
continuum of planets detected between pure water and Earth-like object.
Earth-like planets consist of objects of varying mass but following the
Earth composition (33\% Fe, 66\% MgSiO$_3$ by mass). Wet Super-Earths
consist of an Earth-like core made of 33\% iron and 66\% silicates and a
significant fraction of water outside the core
\citep{Valencia_OConnell_Sasselov06,Thomas_Madhusudhan16}. To test the
validity of our EOS at describing these objects, we show in figure
\ref{fig10} the mass-radius relationship obtained by considering
isothermal models and wet super-Earth planets constituted of 50\% water.
The iron and silicates EOSs used are temperature dependent and, similarly
to the water EOS developed here, are described by free energy functional
forms adjusted directly to \textit{ab initio} calculations
\citep{Bouchet_etal13,Mazevet_etal15}.   

\begin{figure}
   \centering
   \includegraphics[width=\columnwidth]{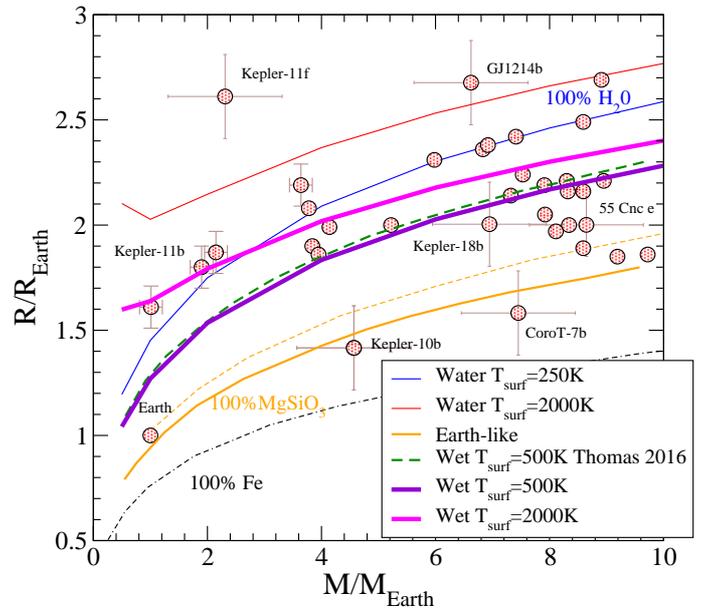}
   \caption{Temperature dependence of the mass-radius relationship for isothermal models of wet super-Earth planets containing 50\% water denoted as Wet. 
   The surface temperatures are indicated in the figure. The dots represent detected planets in this range of planetary mass. \textit{Isothermal models:}  
   pure water (water), silicates (MgSiO$_3$), iron (Fe),  Earth-like composition containing 66\% silicates and 33\% iron (Earth-like) 
   \citep{Bouchet_etal13,Mazevet_etal15}.  The benchmark calculations for wet super-Earth are from \citet{Thomas_Madhusudhan16}. }
              \label{fig9}%
\end{figure}

Figure \ref{fig9} shows the mass-radius relationship for isothermal models, obtained using two different surface temperatures. For a surface temperature of $T_\mathrm{surf}=500$~K, we
obtained a rather satisfactory agreement with the previous calculations of \citet{Thomas_Madhusudhan16}. We point out that the outer
boundary condition used in the current work follows a somewhat different prescription. As mentioned before, we do not consider the vapor phase for surface
temperature below the critical point and use the liquid-vapor boundary at
1~bar as outer boundary conditions. Figure \ref{fig9} shows that the temperature
dependence of the mass-radius relationship is rather significant and can not be neglected when assessing the interior structure of these objects, 
in agreement with the conclusions of \citet{Baraffe_Chabrier_Barman08}.
Indeed, we see in Fig.~\ref{fig9} that several planets lie between the boundaries delimited by the two surface temperatures. As for the case of the
pure water planets, the temperature dependence decreases as the mass of the planet increases and becomes negligible beyond $15M_\mathrm{Earth}$.

\begin{figure}
   \centering
   \includegraphics[width=\columnwidth]{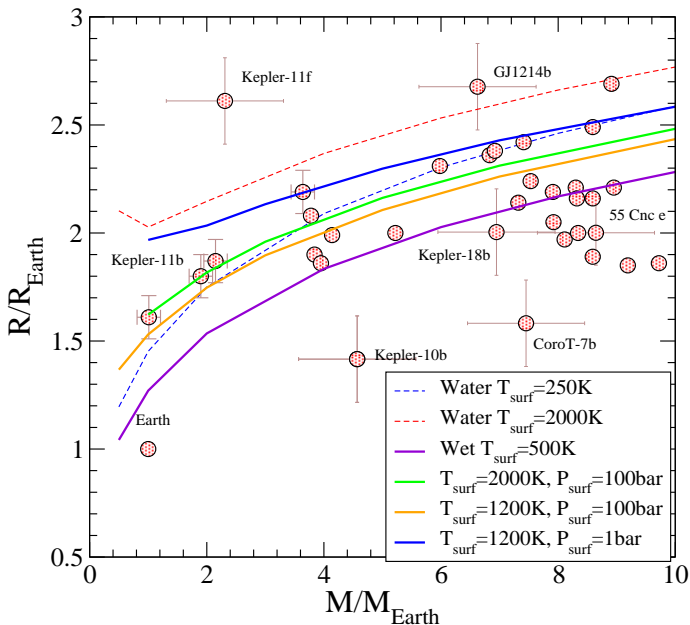}
   \caption{Temperature dependence of the mass-radius relationship for adiabatic models of wet super-Earth planets containing 50\% water as solid lines. 
   The surface temperatures and pressures are indicated in the figure. The dots represent detected planets in this range of planetary mass. 
   The isothermal models for pure water planets (dashed) and wet super Earth (purple) with the corresponding surface temperatures are also displayed in the figure.
   These correspond to the data shown in, respectively, Fig.~\ref{fig8} and Fig.~\ref{fig9}.}
\label{fig10}%
\end{figure}

The wet Earth-like model that consists by 50\% of water is at a midpoint between the dry Earth-like planets and pure water planets. By extension of the
temperature dependence shown for planets composed by 50\% of water, we can anticipate that this also needs to be taken into account to deduce the composition
of objects such as Kepler-18b or 55Cnc that hold a significant fraction of water. In this case, the amount of water deduced will directly depend on the
surface temperature considered. Along the same line of reasoning, we can also anticipate that the overlap between the pure water and wet Earth-like
planet with a surface temperature of $T_\mathrm{surf}=2000$~K shown below $2M_\mathrm{Earth}$ for a composition of 50\% water will expand to planets with larger mass
when the fraction of water is increased.  To make a more quantitative statement on this issue requires to go beyond the internal structure calculations
and to account for an accurate modeling of the atmosphere as well as the amount of light received from the host star. Both these topics are
beyond the scope of this study that aims at validating the EOS for dense water developed here. We will instead turn to the evaluation of the effect of
temperature on the mass-radius relationship beyond the simple isothermal
model where the temperature is kept constant throughout the planet.

The EOS developed in the current work provides the total entropy thanks to
the parametrization of the Helmholtz free energy using the \textit{ab
initio} results. This allows us to calculate more realistic interior
models by considering a water layer undergoing nearly adiabatic convection. 
The uncertainty in defining $S_0$ pointed out in Sect.~\ref{sect:fit} does not
affect these results as long as the convective layer remains within a single thermodynamic phase.
The previous isothermal models and the
adiabatic ones bracket the maximum impact of the water EOS upon the mechanical structure of the body.  We see in
Fig.~\ref{fig10} that the effect of temperature on the mass-radius
relationship is almost two times more important when considering the water
layer as adiabatic rather than isothermal. When considering a surface
pressure of 1 bar, Fig.~\ref{fig10} shows that the radius obtained
up to 10$M_\mathrm{Earth}$ is significantly bigger than the pure water
case at zero temperature. This effect is the most spectacular for surface
temperature above the critical temperature. It remains limited when the
surface temperature is below the critical point. We also point out that
the surface pressure tends to reduce this effect. Figure \ref{fig10} shows
that the radius decreases by 20\% at 1200~K when the surface pressure
increases from 1~bar to 100~bar. This suggests that neglecting the effect
of temperature when identifying the internal structure of exoplanets tends
to over-estimate the overall amount of water, especially for objects close
to their parent star and receiving a significant amount of light. This
needs to be tempered by the probable escape of the atmosphere under these
particular conditions and will need to be assessed on a case by case basis
with a proper treatment of the atmosphere.

\subsection{Core of giant planets}
We now turn to the last situation where a temperature-dependent EOS for dense water is important for planetary modeling, the core of giant planets. Following
the well accepted core-accretion scenario \citep{Pollack_etal96}, a significant amount of water is expected in the core of giant planets. This potentially
significant amount of water stems from the likely composition of the initial core that triggered the accretion of a large fraction of hydrogen and helium.
The exact amount of water as well as the size of the core for a planet like Jupiter is a matter of debate and one of the main scientific goal for the Juno
mission \citep{Bolton_etal17}. The probe is currently measuring Jupiter's high order gravitational moments for Jupiter to decipher the amount of metallic
element present in the interior as well as their distribution throughout the planet \citep{Wahl_etal17}. 
 This goal requires an accurate modeling
of the EOS for all the elements potentially constituting the planet, including water.

\begin{figure}
   \centering
   \includegraphics[width=\columnwidth]{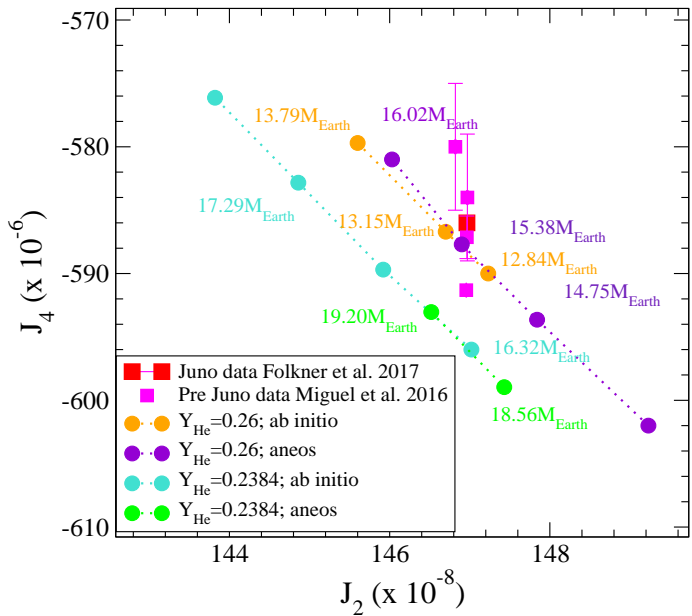}
   \caption{Dependence of Jupiter's first two gravitational moments on the
    dense water EOS used, namely the present and ANEOS ones. 
    The calculations are performed in a two layers model, for a fixed value of
    the mass fraction of He in the envelope, $Y_\mathrm{He}$, indicated in the graph 
    with the mass of the dense water core varying. The size of the pure water core is indicated
    for a few sample points in the figure.  
    The pre-Juno values of the gravitational moments are the values collected in 
    \citet{Miguel_Guillot_Fayon16} while the Juno data are from
    \citet{folkner_2017}.}
              \label{fig11}%
  \end{figure}

Figure \ref{fig11} shows the dependence of Jupiter first two gravitational moments, $J_2$ and $J_4$, on the dense water EOS used in the modeling of the
interior.  Jupiter's interior is obtained by solving the standard
hydrostatic equilibrium equations (\ref{eqhyd1})--(\ref{eqhyd3}) and by considering the
planet as composed of an
H-He envelope and a pure water core (note that we assume a fully adiabatic interior profile). 
The hydrogen and helium EOSs used for the envelope are also based on \textit{ab initio} simulation results
\citep{Caillabet_Mazevet_Loubeyre11,soubiran2012}. 
The gravitational moments are calculated using the theory of figures 
to the third order \citep{planetary_interiors}.

Figure \ref{fig11} shows the values of the first two gravitational moments 
obtained using two different water EOSs, namely the present and the widely used ANEOS ones, to describe the core and by
considering two different helium concentrations. The size of the core is
varied to the values indicated in the graph. The first 
mass fraction of helium, $Y_\mathrm{He}=0.2384$, corresponds to the Galileo measurements,
while a slightly higher mass fraction, $Y_\mathrm{He}=0.26$, allows us to
reproduce the values of the two gravitational moments $J_2$ and $J_4$.
Within a two layers model of Jupiter, the first case reproduces the
observed radius of the planet as well as the value of $J_2$ measured but,
as shown in figure \ref{fig11}, misses the $J_4$ moment by slightly more
than 1\%. Conversely, in the case where the mass fraction of helium is increased to
$Y_\mathrm{He}=0.26$, the observed radius is underestimated by slightly
more than 1\% while matching the first two gravitational moments. This
shortcoming of the two layer model for Jupiter, namely a central core 
surrounded by a homogeneous gaseous envelope, has been well documented
elsewhere \citep{Miguel_Guillot_Fayon16}. We point out here that this
issue is not resolved by using different water EOSs for the core. 

We see in figure \ref{fig11} that using different water EOSs leads to
different predictions regarding the size of the core. 
These two predictions in the size of the core differ by close to 20\%, and do not depend on
the helium concentration. We note that this also translates into different average and maximum densities reached in the central core.
The EOS developed here predicts a pure water core density more than 25\% higher than in the case of the widely used ANEOS one.
The highest density reached in the former case is close to 12.45 g~cm$^{-3}$, while it remains close to 9.67 g~cm$^{-3}$ in the latter case.
This difference is comparable with the variation of the core density when a pure ice core is replaced by a pure rock core, 
as pointed out by \citet{guillot-99}.
In our case, this can be traced back to  significant
discrepancies between the two EOSs, which predict different
compressibilities for pressures and temperatures relevant to Jupiter's
inner core. These conditions correspond to pressures above 40~Mbar and temperatures between
15\,000~K and 20\,000~K depending on the hydrogen-helium EOSs used \cite{Miguel_Guillot_Fayon16}. 
In this thermodynamical region, the two water EOSs predict pressures that differ by more
than 10\%. This shows that the current EOS is useful to calculate accurately the exact size of the core
that is potentially present at the center of giant planets. 

\section{Summary}

In summary, we developed an equation of state for water applicable to the
full range of thermodynamic conditions relevant to planetary modeling.
This encompasses the range from outer layers of wet super-Earth to the
core of giant planets. This equation of state includes an
evaluation of the entropy (within an arbitrary constant value $S_0$,
Sect.~\ref{sect:fit}, which can differ from one phase to the other but 
has no impact on all other thermodynamic quantities) 
thanks to the parametrization of the Helmholtz
free-energy functional form on the \textit{ab initio} results. Using this
EOS, we show that the temperature dependence of the EOS needs to be
accounted for when analyzing the composition and nature of exoplanets
using the standard mass-radius relationship, a point already stressed by
\citet{Baraffe_Chabrier_Barman08}. 
We also show that an accurate description of the
thermodynamic properties of dense water is required to deduce the mass of
the core of giant planets from gravitational moments as it is currently
measured for Jupiter by the Juno mission. As discussed in
Sect.~\ref{sect:fit}, the EOS developed here is not very accurate for the
solid ice phases. This does not noticeably affect the planetary gross
properties, such as the mass-radius relationship, but may produce an
error of several percent in calculations of temperature
profiles. To include explicitly all the ice phases as well as treating 
the super-ionic phase will be the topic of
further development of the present EOS. 

\begin{acknowledgements}
      Part of this work was supported by the SNR grant PLANETLAB 12-BS04-0015 and the Programme National de Planetologie (PNP) of CNRS-INSU co-funded by CNES. Funding and support from Paris Sciences et Lettres (PSL) 
      university through the project origins and conditions for the emergence of life is also acknowledged. This work was performed using HPC resources from GENCI- TGCC (Grant 2017- A0030406113)
\end{acknowledgements}

%   \bibliography{h2o} % your references Yourfile.bib

%%%%%%%%%%%%%%%%%%%%%%%%%%%%%%%%%%%%%%%%%%%%%%%%%%%%%%%%%%%%%%%%%


\end{document}